\documentclass[twocolumn,floatfix,eqsecnum,nofootinbib,aps]{revtex4} 
\usepackage{epsfig}
\usepackage{subfigure}
\usepackage{graphicx}
\usepackage{hyperref}

\def\Tr{{\rm Tr}}

\def\eq#1{(\ref{#1})}

\def\s0#1#2{\mbox{\small{$ \frac{#1}{#2} $}}}
\def\0#1#2{\frac{#1}{#2}}

\def\grgl{\:\hbox to -0.2pt{\lower2.5pt\hbox{$\sim$}\hss}{\raise3pt\hbox{$>$}}\:}
\def\klgl{\:\hbox to -0.2pt{\lower2.5pt\hbox{$\sim$}\hss}{\raise3pt\hbox{$<$}}\:}

\begin{document}

\title{Black holes and asymptotically safe gravity}
\author{Kevin Falls}
\affiliation{ 
\mbox{${}$Department of Physics and Astronomy, University of Sussex,
Brighton BN1 9QH, U.K.}}
\author{Daniel F. Litim}
\affiliation{ 
\mbox{${}$Department of Physics and Astronomy, University of Sussex,
Brighton BN1 9QH, U.K.}}
\author{Aarti Raghuraman\footnote{Present address: Department of Physics, Syracuse University, Syracuse, NY13244,USA.}}
\affiliation{ 
\mbox{${}$Department of Physics and Astronomy, University of Sussex,
Brighton BN1 9QH, U.K.}}
\begin{abstract}
Quantum gravitational corrections to black holes are studied in four
and higher dimensions using a renormalisation group improvement of the
metric.  The quantum effects are worked out in detail
for asymptotically safe gravity, where the short distance physics is
characterized by a non-trivial fixed point of the gravitational
coupling. We find that a weakening of
gravity implies a decrease of the event horizon, and the existence of a 
Planck-size black hole remnant
with vanishing temperature and vanishing heat capacity.
The absence of curvature singularities is generic and discussed together with
the conformal structure and the Penrose diagram of asymptotically safe black holes. 
The production cross section 
of mini-black holes in energetic particle collisions, such as those at the 
Large Hadron Collider, is analysed within low-scale quantum gravity models. 
Quantum gravity corrections imply that cross sections display a threshold, 
are suppressed in the Planckian, and 
reproduce the semi-classical result in the deep trans-Planckian region.
Further implications are discussed.
\end{abstract}
\maketitle


\section {Introduction}

Black holes are intriguing solutions to Einstein's classical equations for gravity,  
characterized by conserved global charges such as total mass, angular momentum, or electric charge.
Most prominently black holes display an event horizon which classically cannot be crossed by light 
rays emitted from their interior. The simplest black hole solution in four dimensions, the Schwarzschild black hole,
 has been discovered  nearly a century ago  \cite{Schwarzschild:1916uq}, and
 many more solutions with increasing degree of complexity are known by now
both in lower and in higher dimensions.  
The latter have received much attention 
recently due to qualitatively new solutions such as black rings 
which cannot be realised in a low dimensional setup \cite{Emparan:2008eg}. 

Recently, the physics of higher dimensional black holes has become 
of particular interest for the phenomenology of
particle physics at colliders. In models where gravity propagates in a higher-dimensional 
space time while Standard Model particles are constraint to a four-dimensional brane 
\cite{ArkaniHamed:1998rs,ArkaniHamed:1998,Randall:1999ee,Randall:1999vf}, 
the fundamental quantum gravity scale  
is as low as the electroweak scale. This open the exciting possibility that particle
colliders such as the LHC could become the first experiment to provide evidence for the 
quantisation of gravity. Signatures of low-scale quantum gravity from particle collisions 
include real and 
virtual graviton effects  \cite{Giudice:1998ck},
and the production  and  decay of TeV size black 
holes  \cite{Dimopoulos:2001hw,Giddings:2001bu}.

It is widely expected that a 
semi-classical description of black hole production and decay is applicable provided curvature
effects remain small, and as long as the black hole mass is large compared to 
the Planck scale \cite{Banks:1999gd,Hsu:2002bd}.
Then the fundamental
black hole production cross section is estimated by the geometric one, 
modulo grey body factors reflecting impact parameter dependences and
inefficiencies in the formation of a horizon \cite{Kanti:2004nr,Webber:2005qa,Giddings:2007nr}. 

The inclusion of quantum gravitational corrections 
to the dynamics of space-time becomes a challenge once the black hole mass  
approaches the fundamental Planck scale. Here, 
quantum gravity effects are  central for  black hole production, decay, 
or the final stages of a gravitational collapse. 
Furthermore, the quantisation of matter fields on a black hole background and the very 
notion of a black hole temperature 
has to be revisited once quantum fluctuations of space-time itself become dominant.
An understanding of the Planckian regime should clarify the 
so-called `information paradox' and the ultimate fate of an evaporating black hole.  

Presently, a complete quantum  gravity description of the above phenomena is 
not at hand. Furthermore, the standard perturbative quantisation for gravity still faces problems. 
Important advances, however,  have been achieved along the lines of Steven Weinberg's 
asymptotic safety scenario  \cite{Weinberg,Weinberg:2009ca,Weinberg:2009bg,Weinberg:2009wa}.  This set-up circumvents the virulent divergences 
encountered within perturbation theory
and leads to well-defined physical observables, such as $S$-matrix elements, 
provided that gravity displays a non-trivial high-energy fixed point 
under the renormalisation group 
\cite{Litim:2006dx,Niedermaier:2006ns,Niedermaier:2006wt,Percacci:2007sz,Litim:2008tt,Litim:2011cp}. 
This intriguing picture implies a non-perturbative ultraviolet completion for gravity, where 
the metric fields remain the fundamental degrees of freedom.  Most importantly, the low 
energy regime of classical general relativity is linked with the high energy 
regime by a well-defined, finite, renormalisation group trajectory \cite{Weinberg,Litim:2008tt,Litim:2011cp}.

In this paper, we  study quantum corrections to black holes in higher
dimensions in the context of asymptotically safe gravity, 
the main results of which have been summarized in \cite{erg2008,MScKevin,MScAarti,FHL}. 
It is our central assumption that the leading quantum 
gravity corrections to black hole  metrics are accounted for by replacing Newton's coupling constant 
through a `running' coupling which evolves under the renormalisation 
group equations for gravity.  Our approach is informed by recent RG results for higher dimensional  quantum gravity 
\cite{Litim:2003vp,Litim:2001up,Litim:2006dx,Fischer:2006fz,Fischer:2006at,Litim:2008tt,Litim:2011cp},
and by earlier black hole studies in the four-dimensional case
\cite{Bonanno:2000ep,Bonanno:2006eu,Bonanno:2009nj}.
Our findings are relevant for the phenomenology of e.g.~mini-black hole production at colliders. 
Further signatures of asymptotically safe quantum gravity at colliders have been analysed in \cite{Litim:2007iu,Litim:2007ee,Hewett:2007st,Gerwick:2009zx,Gerwick:2011jw}.

 The paper is organized as follows. We first recall the essentials of
 classical black holes, and outline the qualitative picture
 (Sec.~\ref{QP}). This is followed by a discussion of the
 renormalisation group equations for the running of Newton's coupling
 within asymptotically safe gravity
 (Sec.~\ref{QG}). We construct improved black
 holes in four and higher dimensions, and analyse their
 main characteristics including the horizon
 structure, mass dependence, the existence of smallest black
 holes  (Sec.~\ref{QBH}), as well as their singularity and causality structure
 (Sec.~\ref{STS}). Our findings are applied to the physics of black
 hole production in higher dimensional scenarios with low-scale
 quantum gravity (Sec.~\ref{production}). We close with a discussion
 of the main results and indicate further implications
 (Sec.~\ref{Discussion}).

\section{Generalities}\label{QP}
In this section, we recall the basics of classical black holes,
introduce some notation, outline the renormalisation group improvement
for black hole metrics and discuss first implications.

\subsection{Schwarzschild metric}
The classical, static, spherically symmetric, non-charged black hole
solution to Einstein's equation is the well-known Schwarzschild black
hole \cite{Schwarzschild:1916uq}.  Its line element in $d\ge 4$ dimensions is
given by \cite{Tangherlini:1963bw} (see also \cite{Myers:1986un})
\begin{equation}\label{ds2} 
ds^2=-f(r)\ dt^2 +\frac{dr^2}{f(r)} + r^2\ d\Omega^2_{d-2}\,.
\end{equation}
The lapse function
\begin{equation}\label{f}
f(r)=
1-\frac{G_N\,M}{r^{d-3}}
\end{equation}
depends on Newton's
coupling constant $G_N$  in $d$ dimensions.  The reduced  black hole mass $M$ is
\begin{eqnarray} \label{M-def}
M=\frac{8\,\Gamma(\frac{d-1}{2})}{(d-2)\pi^{(d-3)/2}}\, M_{\rm phys}\,,
\end{eqnarray} 
with $M_{\rm phys}$ the physical mass of the black hole.
In terms of these, the classical Schwarzschild radius $r_{\rm cl}$ is given as
\begin{eqnarray} \label{classicalBH-def}
r_{\rm cl}^{d-3}&=& G_N \,  M\,.
\end{eqnarray} 
The black hole solution is continuous in the mass parameter $M$ and
displays a Bekenstein-Hawking temperature inversely proportional to
its mass. For large radial distance $r\to\infty$, we observe $f(r)\to
1$, indicating that the geometry of a Schwarzschild space-time becomes
flat Minkowskian.  The coordinate singularity at $r=r_{\rm cl}$ where
$f(r_{\rm cl})$ vanishes, defines the event horizon of the black hole.
In the short distance limit $r\to 0$ we observe a divergence in
$f(r)$, reflecting a metric and curvature singularity at the origin.

\begin{center}
\begin{table*}[t]
\begin{tabular}{c|c|c|c|c}
case 
& short distance index
& gravity 
& horizons 
&${}$\quad  $f(r\to 0)$${}$\quad${}$  \\
\hline
(i)
& ${}$\quad  $\alpha<d-3$ \quad ${}$ 
& ${}$\quad strong, if $\alpha<0$;\quad weak, if $\alpha>0$${}$\quad ${}$ 
& one      
& ${}$\quad singular${}$\quad\\[.5ex] 
(ii)
&${}$\quad $\alpha=d-3$\quad ${}$
& weak
& none, one or more  
& finite\\[.5ex]
${}$\quad${}$ (iii)\quad${}$
&$\alpha>d-3$
& weak
& none, one or more  
&1\\ 
\end{tabular}
\caption{\label{T1} Horizons of quantum-corrected Schwarzschild black holes assuming a
  scale-dependent gravitational coupling strength \eq{G(r)1} at short distances for various
  dimensions and in dependence on the short distance index $\alpha$ (see text).}
\end{table*}
\end{center}
\subsection{Improved metric}\label{IM}
The classical black hole is modified once quantum gravitational
effects are taken into account. In general, quantum fluctuations will
modify the gravitational force law by turning Newton's coupling $G_N$
into a distance-dependent ``running" coupling $G(r)$,
\begin{equation}
G_N\to G(r)\,.\label{RGimproved}
\end{equation}
It is the central assumption of this paper that the leading quantum
gravitational corrections to the black hole are captured by the
replacement \eq{RGimproved} in the metric \eq{f}. This
``renormalisation group improvement" should provide a good description
of the leading quantum corrections, because the primary, explicit,
dependence of the Schwarzschild black hole on the gravitational sector
is only via Newton's coupling $G_N$. Furthermore, the classical black
hole solution is continuous in its mass parameter $M$, and the effects
of quantum corrections are parametrically suppressed for large black
hole mass with $M_D/M$ serving as an external, small, control
parameter.
Whether gravity becomes ``strong" at shortest distances, or ``weak",
will depend on the ultraviolet completion for gravity and the related
running under the renormalisation group.  

Next we discuss the main
implications arising from a running gravitational
coupling. For the sake of the argument, we parametrize $G(r)$ as
\begin{equation}\label{G(r)1}
G(r)=r_{\rm char}^{d-2} \left(\frac{r}{r_{\rm char}}\right)^{\alpha}
\end{equation}
for sufficiently small $r$, where $r_{\rm char}$ denotes a characteristic
length scale where quantum corrections become dominant. The index
$\alpha$ then parametrizes the gravitational coupling strength at
short distances, with $\alpha>0$ $(\alpha<0)$ denoting a decrease
(increase) of $G(r)/G_N$ at small distances, respectively, and the
classical limit $\alpha=0$ where $r_{\rm char}$ is given by the Planck
length $r_{\rm char}=1/M_D$.  The behaviour of $f(r\to 0)$, and the
solutions to the horizon condition $f(r)=0$ then teach us how the
RG-improved black hole depends on the quantum effects parametrized by
$\alpha$. The qualitative pattern is summarised in Tab.~\ref{T1}.  We distinguish
three cases, depending on the short distance index $\alpha$:
\begin{itemize}
\item[(i)] $\alpha<d-3$. In this case the gravitational coupling
  either increases with decreasing $r$, or even decreases slightly,
  though not strongly enough to overcome the enhancement due to the
  $\s01{r^{d-3}}$-factor in \eq{f}. Therefore $f(r)$ unavoidably has
  to change sign leading to a horizon. This includes the classical
  case $\alpha=0$, and all cases of strong gravity corresponding to a
  diverging $G(r)/G_N$ for small $r$. Interestingly, even if gravity
  weakens at short distances with an index $0<\alpha<d-3$, we still
  observe a horizon for arbitrary small black hole masses.
\item[(ii)] $\alpha=d-3$. In this case, we have a finite limit $f(r\to
  0)=f_0$. For $f_0<0$, this necessarily enforces a horizon, similar
  to case (i). For $f_0>0$, the situation is analogous to case (iii).
\item[(iii)] $\alpha>d-3$. In this case, $G(r)$ weakens fast enough to
  overcome the enhancement due to $\s01{r^{d-3}}$. Therefore $f(r\to
  0)\to 1$ and $f(r)$ may display either several zeros, a single one,
  or none at all, leading to several, one or no horizon depending on
  the black hole mass $M$ and the precise short-distance behaviour of
  $G(r)$.
\end{itemize}
We conclude that for $\alpha >d-3$ the Schwarzschild black hole may no
longer display a horizon for all mass, whereas for $\alpha<d-3$ a
horizon is guaranteed for all $M$.
Which of these scenarios is realised depends on the short-distance
behaviour of gravity.
In the remaining part of the paper we access this picture
quantitatively, using the renormalisation group for gravity.

\section{Asymptotically safe gravity} \label{QG}

In this section, we discuss field theory based approaches to quantum 
gravity including effective theory and the asymptotic safety scenario for gravity, 
and provide the renormalisation group running for Newton's coupling.

\subsection{Effective theory for gravity}\label{ET}
In the absence of a complete theory for quantum gravity, quantum corrections  
of the form \eq{RGimproved} can be accessed in the weak gravity regime 
using methods from effective theory \cite{Donoghue:1993eb,Burgess:2003jk}. 
In practice, this amounts to an ultraviolet regularisation of the theory by an 
UV cutoff $\Lambda$ of the order 
of the fundamental Planck scale. In the weak gravity regime where 
$r\, M_D\gg 1$ with Planck mass $M_D=(G_N)^{-1/(d-2)}$, it has been
found that
\begin{equation}
{G(r)}={G_N}\left(1-\frac{\omega\,G_N}{r^2}\right)
\end{equation}
in four dimensions, and at the one-loop order 
\cite{Hamber:1995cq,Bjerrum-Bohr:2002ks,Bjerrum-Bohr:2002kt,Akhundov:2006gh}, 
with $\omega>0$ (see \cite{Duff:1974ud} for earlier results). In higher dimensions, no effective theory 
results are available and 
thus we have to  provide the relevant RG input  from a different source.
 
\subsection{Asymptotic safety}\label{AS}
Asymptotic safety of gravity is a scenario where gravity exists as a well-defined
fundamental local quantum theory of the metric field \cite{Weinberg,Weinberg:2009ca,Weinberg:2009bg,Weinberg:2009wa}. This set-up goes one step 
beyond an effective theory approach: it assumes that the 
ultraviolet cutoff $\Lambda$ from effective theory can safely be removed, 
$\Lambda\to\infty$, whereby the relevant gravitational couplings approach 
a non-trivial fixed point \cite{Litim:2008tt,Litim:2006dx,Niedermaier:2006ns,Niedermaier:2006wt,Percacci:2007sz,Litim:2011cp}.
We briefly recall the main picture. Consider
the dimensionless gravitational coupling 
\begin{equation}
g(\mu)=G(\mu)\mu^{d-2}\equiv G_0 Z_G^{-1}(\mu)\mu^{d-2}\,,
\end{equation} 
where $\mu$ denotes the RG momentum scale,
$G(\mu)$ is the running Newton coupling, 
and $Z_G$ the gravitational wave function factor.
The wave function factor is normalised as
$Z_G(\mu_0)=1$ at some reference scale $\mu_0$ with $G(\mu_0)$ given
by Newton's constant $G_0\equiv G_N$.
Then the gravitational Callan-Symanzik equation reads \cite{Litim:2003vp,Litim:2001up,Litim:2006dx,Niedermaier:2006ns}
\begin{equation}\label{dg}
\frac{{\rm d}g(\mu)}{{\rm d}\ln \mu} = (d-2+\eta)g(\mu)\,.
\end{equation}
Here $\eta=-\mu\partial_\mu \ln Z_G$ denotes the anomalous dimension
of the graviton.   In general, the anomalous
dimension depends on all couplings of the theory. Due to its
structure, \eq{dg} can achieve two types of fixed points. At small
coupling, the anomalous dimension vanishes and $g=0$ corresponds to
the non-interacting ($i.e.$~Gaussian) fixed point of \eq{dg}. This
fixed point dominates the deep infrared region of gravity
$\mu\to 0$. In turn, an interacting fixed point $g_*$ is achieved
if the anomalous dimension of the graviton becomes non-perturbatively
large,
\begin{equation}\label{eta}
\eta_* = 2-d\,.
\end{equation} 
A non-trivial fixed point of quantum gravity in $d>2$ implies a
negative integer value for the graviton anomalous dimension,
counter-balancing the canonical dimension of $G$.  As a consequence, 
$G(\mu)\to
g_*/\mu^{d-2}$ in the vicinity of a non-trivial fixed point. In the UV
limit where $\mu\to \infty$, the gravitational coupling $G(\mu)$
becomes arbitrarily weak.

\subsection{Renormalisation group}\label{RG}
The above picture is  substantiated through explicit renormalisation group studies for 
gravity. A powerful tool is given by the functional renormalisation group \cite{ERG-Reviews,ERG-Wetterich,ERG-Polonyi,Pawlowski:2005xe,Gies:2006wv}, 
based on the idea  of integrating-out momentum modes from a path-integral 
representation of quantum field theory \cite{Litim:2006dx,Niedermaier:2006ns,Niedermaier:2006wt,Litim:2008tt,Litim:2011cp}. 
The corresponding `flowing' effective action for gravity reads
\begin{equation}\label{EHk}
 \Gamma_k=
\0{1}{16\pi G_k}\int d^dx \sqrt{g}\left[-R(g)+\cdots\right]\,.
\end{equation}
Here  $R(g)$ denotes the Ricci scalar, and the dots in \eq{EHk}
stand for the cosmological constant, higher dimensional operators in
the metric field, gravity-matter interactions, a classical gauge
fixing and ghost terms.  Furthermore, all couplings in \eq{EHk}  
are `running' couplings as functions of the Wilsonian momentum scale
$k$, which now takes over the role of the RG scale $\mu$ discussed 
above.  The action \eq{EHk} reduces to the standard quantum effective action
in the limit $k\to 0$, where all quantum fluctuations are integrated out.
For $k\ll M_D$, the gravitational sector is well-approximated by
the Einstein-Hilbert action with $G_k\approx G_0$, and similarly for
the gravity-matter couplings.  The corresponding operators scale
canonically.  At $k\approx M_D$ and above, the non-trivial RG running
of gravitational couplings becomes important.

An exact functional flow equation which governs the $k$-dependence for an action \eq{EHk} 
has been put forward by  Wetterich \cite{Wetterich:1992yh},
\begin{equation}\label{ERG} 
\partial_t \Gamma_k=
\frac{1}{2} \Tr \left({\Gamma_k^{(2)}+R_k}\right)^{-1}\partial_t R_k
\end{equation} 
and $t=\ln k$. The trace stands for a momentum integration and a sum
over indices and fields, and $R_k(q^2)$ denotes an appropriate
infrared cutoff function at momentum scale $q^2\approx k^2$
\cite{Litim:2000ci,Litim:2001fd}.  The flow \eq{ERG} can be seen as a functional Callan-Symanzik equation, where the mass term is replaced by a momentum-depedent mass $k^2\to R_k(q^2)$  \cite{Symanzik:1970rt}. When implemented for gravity \cite{Reuter:1996cp},
an additional background field is introduced to achieve diffeomorphism
invariance within the background field technique \cite{ERG-Reviews,ERG-Wetterich,ERG-Polonyi,Pawlowski:2005xe,Gies:2006wv,Freire:2000bq,Reuter:1996cp,Litim:2002ce}.
By now,  a large number of papers have shown the existence
of a non-trivial fixed point for gravity including renormalisation group studies in four and higher dimensions
\cite{Reuter:1996cp,Souma:1999at,Lauscher:2001ya,Lauscher:2002mb,Reuter:2001ag,Percacci:2002ie,Litim:2003vp,Litim:2001up,Bonanno:2004sy,Percacci:2005wu,Fischer:2006fz,Fischer:2006at,Codello:2006in,Codello:2007bd,Machado:2007ea,Rahmede:2009,Benedetti:2009gn,Niedermaier:2009zz,Niedermaier:2011zz,Percacci:2009dt},
 and numerical simulations on the lattice
\cite{Hamber:1999nu,Hamber:2004ew,Hamber:2005vc,Ambjorn:2004qm,Ambjorn:2005db}.

Analytical results for the running of the gravitational coupling have been given in \cite{Litim:2003vp,Litim:2001up}, where
\eq{EHk} has been approximated by the Ricci scalar. The central result is not altered through 
the inclusion of a cosmological constant \cite{Fischer:2006fz}. Using \eq{EHk} and \eq{ERG},
one finds 
\begin{eqnarray}\label{betag0}
\beta_g&=&\0{(1-4dg/c_d)(d-2)g}{1-(2d-4)g/c_d}
\end{eqnarray}
with parameter
$c_d=(4\pi)^{d/2-1}\Gamma(\s0d2+2)$. The scale-dependence of the anomalous dimension 
is given via the scale-dependence of the running coupling,
\begin{eqnarray}
\eta&=&\0{2(d-2)(d+2)\,g/c_d}{2(d-2)\,g/c_d-1}\,.
\end{eqnarray}
We observe a Gaussian fixed
point at $g_*=0$ and a non-Gaussian one at $g_*=c_d/(4d)$.
Integrating the flow \eq{betag0} gives an implicit equation for $G_k$,
\begin{equation}\label{g0-explicit}
\0{G(k)}{G(k_0)}
=\left(\0{g_*-G(k)\,k^{d-2}}{g_*-\,G(k_0)\,k_0^{d-2}}\right)^{(d-2)/\theta}
\end{equation}
with boundary condition $G(k_0) k_0^{d-2}< g_*$, and the non-perturbative scaling exponent $\theta=2d\ 
\0{d-2}{d+2}$.  The fixed point value and the scaling exponent depend slightly on the underlying momentum cutoff \cite{Litim:2003vp,Litim:2001up,Fischer:2006fz}.  Inserting the running coupling
\eq{g0-explicit} into \eq{betag0} shows that the anomalous dimension displays
a smooth cross-over between the IR domain $k\ll M_D$ where $\eta \approx 0$
and the UV domain $k\gg M_D$ where $\eta \approx2-d$.
The cross-over regime becomes narrower with increasing
dimension.   For our purposes, it will be sufficient to approximate the non-perturbative solution \eq{g0-explicit} further by  setting the scaling index $\theta$  to $\theta=d-2$. In the limit where $G(k_0)k^{d-2}_0\ll 1$, we find
\begin{equation}\label{G(k)}
\frac{1}{G(k)}=\frac{1}{G_0} +\omega \,k^{d-2}
\end{equation}
where $ \omega=1/g_*$ is a positive constant, and $G_0=G(k_0=0)$. Note that \eq{G(k)} looks, formally, like a 1-loop equation. The difference here is that the coefficient $\omega$, in general, also encodes information about the underlying fixed point and may be numerically different from the 1-loop coefficient.  This equation captures the main 
cross-over behaviour. 

\begin{figure*} 
\subfigure[\label{disdim}\ Proper and linear distance in various dimensions.]{
\includegraphics[width=.45\hsize]{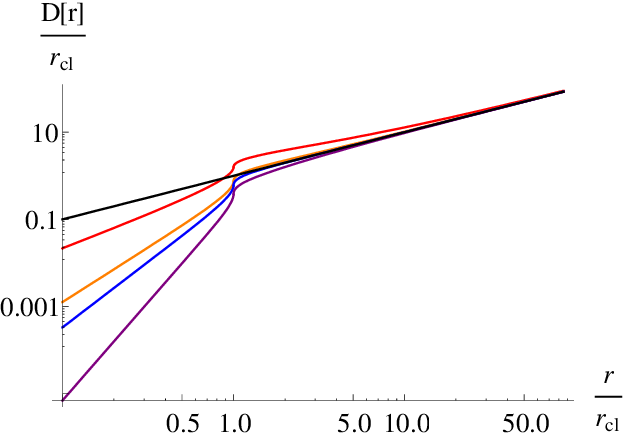}}
\subfigure[\label{LogDis}\ Several distance functions in $d=7$ dimensions.]{
\includegraphics[width=.45\hsize]{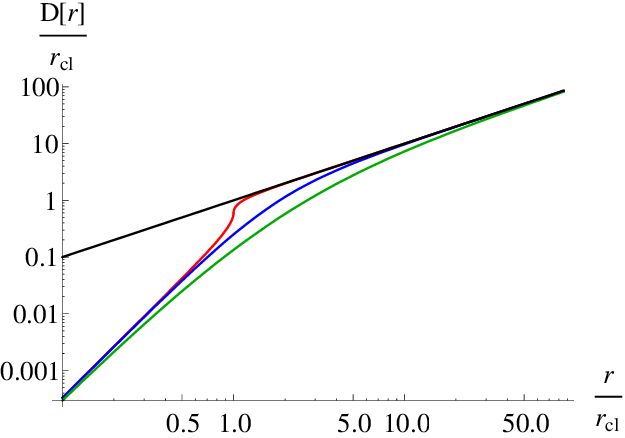}}
\caption{\label{LogDis-compare} Comparison of various distance functions $D(r)$ as functions of $r/r_{\rm cl}$. (a) Proper distance in  $d=4,6,7$ and $10$ dimensions (top to bottom) and  linear matching (straight line). (b) Interpolating expressions \eq{D-interpol} and \eq{D-interpol2}, proper distance matching \eq{distance}, and  linear matching \eq{linear} (bottom to top) in 7 dimension. }
\end{figure*}

\subsection{Relevant scales}\label{Scales}

In order to implement quantum corrections to the classical Schwarzschild 
black hole geometry, we replace the classical coupling $G$ by an $r$-dependent running coupling $G(r)$ under the RG flow. The renormalisation group provides us with a momentum-scale dependent $G(k)$. This requires, additionally, a scale identification between the momentum scale $k$ and the coordinate variable $r$ of the form
\begin{equation}
\label{match}
k(r)=\xi/D(r)\,,
\end{equation}
such that \begin{equation} \label{G(r)2}
\frac{1}{G(r)}=\frac{1}{G_0}+\frac{\omega\,\xi^{d-2}}{D^{d-2}(r)}\,.
\end{equation} The distance function 
$D(r)$ should be an appropriately chosen length scale which
may depend on other parameters such as eg.~the black hole mass $M$. In
general, the matching coefficient $\xi$ is non-universal and its
numerical value will depend on the RG scheme used to obtain the RG
running of $G(k)$. In a fixed RG scheme, and for a given choice for
$D(r)$, $\xi$ can be computed explicitly using methods discussed in
eg.~\cite{Donoghue:1993eb}.

Next we introduce a variety of distance functions motivated by the
Schwarzschild metric, flat space metric, dimensional analysis, and
interpolations.

$\bullet$  {\it Dimensional analysis.}  The gravitational force on a
test particle in a Schwarzschild space-time depends on two independent
dimensionful parameters, the horizon $r_{\rm cl}$ (or the black hole
mass, respectively) and the radial distance scale $r$. Therefore,
dimensional analysis suggests that a general length scale $D(r)$ can
be written as
\begin{equation}
\label{nonlinear}
D_{\rm da}(r)= c_{\gamma} r^{\gamma}\, r_{\rm cl}^{-\gamma+1}\,.
\end{equation}
for some $\gamma$, and $c_{\gamma}$ a positive constant.  Moreover,
$\gamma$ may depend on dimensionless ratios such as $r/r_{\rm cl}$. An
ansatz taking into account the flat-space limit for $r\to \infty$, and
the deep Schwarzschild regime $r\ll r_{\rm cl}$, is given by
\begin{equation}
\label{da}
D_{\rm da}(r)\propto \left\{
\begin{array}{cc} r & {\rm for}\ \ r>r_{\rm cl}\\
 r^{\gamma}\, r_{\rm cl}^{-\gamma+1} & {\rm for}\ \ r<r_{\rm cl}
 \end{array}
 \right. 
\end{equation}
with coefficient $\gamma$. In the parametrisation \eq{G(r)1}, this
corresponds to the short-distance index $\alpha=\gamma(d-2)$. For
$\gamma>1$ $(\gamma<1)$, the matching enhances (counteracts) the RG
running of \eq{G(k)}.  We note that $1/\gamma\to 0$ corresponds to a
decoupling limit where gravity is switched-off at scales below $r_{\rm cl}$.

$\bullet$  {\it Proper distance.} A different matching is obtained by
identifying the RG momentum scale $k$ with the inverse diffeomorphism
invariant distance $D_{\rm diff}(r)^{-1}$ \cite{Bonanno:2000ep}. Such
a distance is defined through the line integral
\begin{equation}\label{Ddef} 
D_{\rm diff}(r)=\int_C \sqrt{|ds^2|} \,,
\end{equation}
where $C$ is an appropriately chosen curve in spacetime.  Using the
classical Schwarzschild metric, we consider a path $C$ running
radially from $0$ to $r$, thereby connecting time-like with space-like
regions.  With $dt=d\Omega=0$ this defines the proper distance
\begin{equation}\label{distance}
  D_{\rm Schw}(r)=\int_{0}^{r} dr 
\left|1-\left(\frac{r_{\rm cl}}{r}\right)^{d-3}\right|^{-1/2}\,.  
\end{equation}
For any $d$, \eq{distance} has an integrable pole $\sim
1/\sqrt{r-r_{\rm cl}}$ at the connection point between space-like and
time-like regions $r=r_{\rm cl}$.  Analytical expressions for $D_{\rm
  Schw}(r)$ are obtained from \eq{distance} for fixed dimension. We
note that \eq{distance} corresponds to \eq{nonlinear} with an
$(r/r_{\rm cl})$-dependent index $\gamma$. For large $r\gg r_{\rm
  cl}$, we have $D_{\rm Schw}(r) \rightarrow r$, where the
Schwarzschild metric becomes flat corresponding to $\gamma=1$ in
\eq{nonlinear}. For small $r$ \eq{distance} corresponds to
\eq{nonlinear} with $\gamma=(d-1)/2$.

$\bullet$  {\it IR matching.}
If the black hole mass $M$ is sufficiently large compared to the
Planck mass,  we can assume that the only RG relevant length scale in the problem
is given by $r$. Therefore, $r$ is directly
identified with the (inverse) RG scale $k$  \cite{Bonanno:2000ep},
\begin{equation}
\label{linear}
D_{\rm ir}(r)=r\,.
\end{equation}
This matching \eq{linear} corresponds to \eq{nonlinear}
with $\gamma=c_{\gamma}=1$. In the parametrisation \eq{G(r)1}, the short distance behaviour corresponds to
$\alpha=d-2$. We therefore expect the  matching
\eq{linear} to capture the leading quantum effects correctly.

$\bullet$  {\it UV matching.} 
For small $r\ll r_{\rm cl}$, the proper distance $D_{\rm Schw}(r)$ scales like a power-law in $r$. We find
\begin{equation}\label{D-UV}
D_{\rm uv}(r)=\frac{2\,r^{(d-1)/2}}{(d-1)\,r_{\rm cl}^{(d-3)/2}}\,.
\end{equation}
Matching the RG momentum scale with the inverse proper distance \eq{D-UV} leads to
\eq{nonlinear} with $\gamma=c_\gamma^{-1}=(d-1)/2$. In the parametrisation \eq{G(r)1}, this
corresponds to the short-distance index $\alpha=(d-1)(d-2)/2>0$, which for all
$d>3$ is larger than the index $(d-2)$ obtained from linear matching. \\

$\bullet$  {\it Interpolations.} 
For the
subsequent analysis, it is useful to have approximate expressions for $D_{\rm Schw}(r)$ 
\eq{distance} which interpolate properly between \eq{linear} and \eq{D-UV}. 
We use a simple interpolation formula for general
dimension to implement the non-linear matching \eq{distance} into \eq{G(k)}
and write
\begin{eqnarray}\label{D-interpol}
D_{\rm int1}(r)&=&\frac{2\,r^{(d-1)/2}}{(d-1)\,(r_{\rm cl}+\epsilon\,r)^{(d-3)/2}}\\
\label{nonepsilon}
\epsilon&=&\left(\s0d2-\s012\right)^{-2/(d-3)}
\end{eqnarray}
which is exact for $r\to \infty$ and $r\to 0$, and $\epsilon \in [\s049,1]$
for $d \in [4,\infty]$.  Alternatively, we also use
\begin{eqnarray}\label{D-interpol2}
D_{\rm int2}(r)&=&\frac{r}{1+\s012(d-1)\,(r_{\rm cl}/r)^{(d-3)/2}}\,.
\end{eqnarray}

In Fig.~\ref{LogDis-compare} we compare different
distance functions. In Fig.~\ref{disdim}, the
functions \eq{distance} are compared with the linear matching
\eq{linear} in various dimensions.  For large $r$ the proper distance
\eq{distance} approaches $r$ for all $d \geq 4$. As $r/r_{\rm cl}
\searrow1$ we observe that the gradient steepens rapidly due to the
presence of an integrable pole $~ 1/ \sqrt{r-r_{\rm cl}}$. For $r \ll
r_{\rm cl}$ the gradients of each curve are steeper with increasing
$d$ due to an additional dimensional suppression in \eq{distance}. In
Fig.~\ref{LogDis} we fix $d=7$ and observe that for large $r$ the
matchings \eq{distance}. \eq{D-interpol} and \eq{D-interpol2} approach
the correct IR behaviour \eq{linear}. For small $r$ these matchings
approach the UV matching \eq{D-UV}.  We also observe that the rapid
steepening of \eq{distance} around $r/r_{\rm cl}$ implies that the
transition between IR and UV behaviour is well approximated by \eq{da}.

Finally, we provide a link with the  discussion of
Tab.~\ref{T1}, see Sec.~\ref{IM}.  For the distance functions motivated by the
Schwarzschild metric \eq{distance}, \eq{D-UV}, \eq{D-interpol} and
\eq{D-interpol2}, we find the index $\alpha= \frac{1}{2}
(d-1)(d-2)>(d-3)$ for all $d \geq 4$, corresponding to case (iii).  In
the same vein, for $D(r)$ motivated by the flat space metric
\eq{linear} we find $\alpha=d-2$ equally corresponding to case (iii).
Finally, the distance function motivated by dimensional analysis \eq{da} contains a free
parameter $\gamma$, whose natural value is of order one. It leads to
the index $\alpha=\gamma(d-2)$ and hence relates to case (iii) for all
$\gamma>\gamma_c$, where
\begin{equation}\label{gammac}
\gamma_c=\frac{d-3}{d-2}\,.
\end{equation} 
We conclude that for all physically motivated distance functions 
we are lead to
the scenario described by case (iii) in Tab.~\ref{T1}, independently
of the scale identification $k=k(r)$. This, therefore, appears to be a
robust prediction from the renormalisation group running implied
within asymptotically safe gravity.

\begin{figure*}
\subfigure[\label{6dlin}  \  RG running with \eq{f-L} and $\gamma=1$ in $6$  dimensions.]
{\includegraphics[width=.45\hsize]{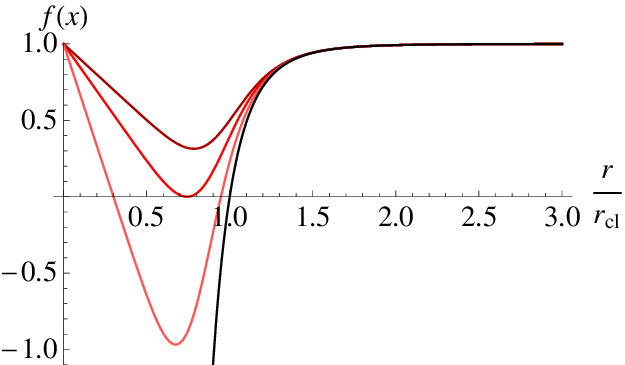}}
\hskip0.09\hsize
\subfigure[\label{nonlin7}  \ RG running with \eq{f-non} and $\gamma=3$ in $7$ dimensions.]
{\includegraphics[width=.45\hsize]{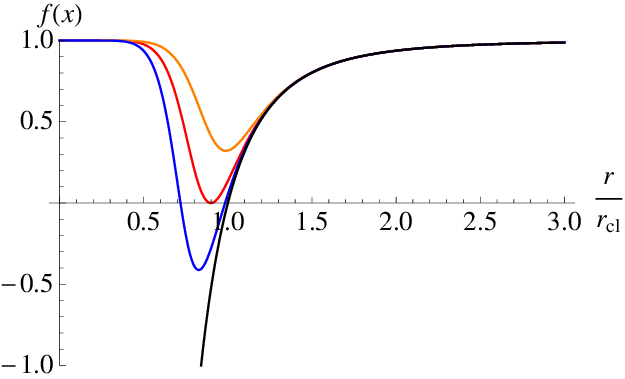}}
\caption{\label{RG-f-compare} Mass and renormalisation group dependence of the
RG improved function $f(x)$ with $x=r/r_{\rm cl}$ in higher dimensions.
From top to bottom: absence of horizon $\Omega > \Omega_c$, critical black hole $\Omega =
  \Omega_c$, semi-classical black hole $\Omega < \Omega_c$, and classical black hole $\Omega=0$.  }
\end{figure*}

\section{Asymptotically safe black holes}\label{QBH}

In this section, we implement the renormalisation group improvement and analyse the
resulting black holes, their horizon structure, and critical
Planck-size mini-black holes.  The asymptotically safe black hole is obtained
by replacing $G_N$ with the running $G(r)$ \eq{G(r)2} in
\eq{classicalBH-def} and \eq{f}, leading to the improved, asymptotically safe, 
lapse function
\begin{equation}\label{f-RG}
f(r,M)=1-G(r,M)\, \frac{M}{r^{d-3}} \,.
\end{equation}
At this point we make two observations.  The improved Schwarzschild
radius $r_s(M)$ is given by the implicit solution of
\begin{equation}
\label{rs_implicit} r_s^{d-3}(M)=M\,G(r_s(M),M)\,.  
\end{equation}
If
\eq{f-RG} has a solution $f(r_s(M),M)=0$, then it follows that the
quantum-corrected horizon is smaller than the classical one
$r_{s}(M)<r_{\rm cl}(M)$. This is a direct consequence of
${G(r,M)}/{G_N}\le 1$ for all $r$. Secondly, if ${G(r,M)}/{G_N}$
decreases too rapidly as a function of $r$, $f(r_s(M),M)=0$ will no
longer have a real solution $r_s(M)\ge 0$, implying the absence of a
horizon.

\subsection{Horizons}\label{horizons}

To see the above picture quantitatively, we analyse the horizon
condition analytically, also comparing various matching conditions.
For a general matching the dimensional analysis \eq{nonlinear} leads
to a running Newton's constant \eq{G(r)2} of the form 
\begin{equation}
\frac{G_0}{G(r)}= 1+\frac{\tilde{\omega}\,G_{0}}{r_{\rm
    cl}^{d-2}}\left(\frac{r_{\rm cl}}{r}\right)^{\gamma(d-2)} 
\end{equation}
with $r_{\rm cl}$ as in \eq{classicalBH-def} and \begin{equation}
\tilde{\omega}=\omega(\xi/c_\gamma)^{d-2} \end{equation} This leads to a lapse
function given by 
\begin{equation} \label{gen-f}
f(x)=1-\frac{1}{x^{d-3}}\frac{x^{\gamma(d-2)}}{x^{\gamma(d-2)}+\Omega}\,,
\end{equation} 
where we have also introduced the variables
\begin{eqnarray}\label{sol-RG-2}
x&=&r/r_{\rm cl}\\
r_{\rm cl}&=&( M\,G_0)^{1/(d-3)}\\
\label{sol-RG-3}
\Omega&=&\tilde\omega\, (M_D/ M)^{\frac{d-2}{d-3}}\,.
\end{eqnarray} 
We define the $d$-dimensional Planck mass as $G_0=M_D^{2-d}$
corresponding to the convention used by Dimopoulos and Landsberg
\cite{Dimopoulos:2001hw}.  The parameter $\Omega$ captures the RG
running of Newton's coupling, and the mass and matching parameter
dependence. The classical black hole corresponds to $\Omega=0$ which
is achieved in the limit of vanishing quantum corrections $\omega\to
0$ or in the limit of infinite black hole mass $M\to\infty$.
Therefore, the horizon condition $f(x)=0$ always includes the
classical solution $x=1$ at $r=r_{\rm cl}$ for $\Omega=0$.

\begin{figure*}
\subfigure[\label{gammaOmega} \ Horizons $r_s$ as a function of
  $\Omega/\Omega_c$ for $d=7$. 
  Upper/ lower branch are event/ Cauchy
  horizon]{\includegraphics[width=.45 \hsize]{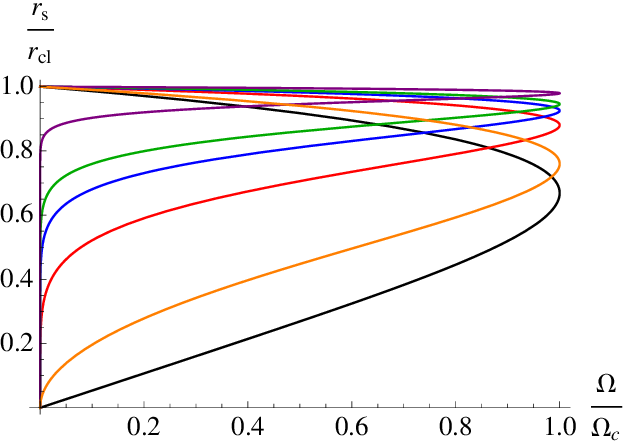}}
\hskip.09\hsize
\subfigure[\label{gammaf} Shape of $f(x)$ with $x=r/r_{\rm cl}$ at criticality $\Omega=\Omega_c$.
The minima denote the degenerate horizon $x_c$.
  ]{\includegraphics[width=.45 \hsize]{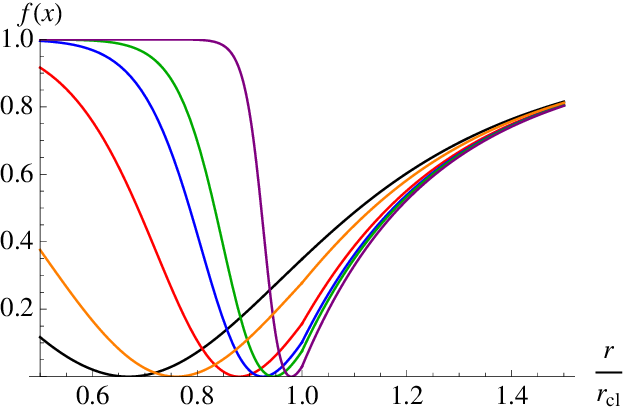}}
\caption{\label{gamma-dependence}Dependence of Cauchy and event horizon, and of the metric coefficient $f(x)$ at criticality  $\Omega=\Omega_c$, on the parameter  $\gamma=1,\s065,2,3,4$, and $10$ [panel (a): from bottom to top, panel (b): from left to right] in $d=7$ dimensions. Metric singularities are absent for $\gamma\ge\gamma_{\rm dS}=\frac{d-1}{d-2}$. Large values of $\gamma\to\infty$  represent the decoupling limit (see text).}
\end{figure*}

For simplicity, we begin with the case $\gamma=1$ corresponding to the
IR matching \eq{linear}, where $f(x)$ takes the form
 \begin{equation}\label{f-L}
f(x)=1-\frac{x}{x^{d-2}+\Omega}\,.
\end{equation}
For $\Omega>0$, the horizon condition becomes
\begin{equation}\label{sol-RG-1}
0=x^{d-3}-1+\Omega/x\,.
\end{equation}
We find three qualitatively different solutions, depending on the
value of $\Omega$ (see Fig.~\ref{6dlin}). In general, \eq{sol-RG-1}
has $d-2$ possibly complex roots $x(\Omega)$. For sufficiently small
$\Omega$, two of these are positive real with
$0<x_-(\Omega)<x_+(\Omega)\le 1$ and correspond to a Cauchy horizon
$x_-\equiv r_w/r_{\rm cl}$ and an outer horizon $x_+\equiv r_s/r_{\rm
  cl}$. In even dimensions, the remaining roots are complex congugate
pairs, whereas in odd dimensions, one of the remaining roots is real
and negative. Analytical solutions are obtained for $x_\pm(\Omega)$ as
a power series in $\Omega$ for any $d>3$. With increasing $\Omega$
(decreasing black hole mass $M$), real solutions to \eq{sol-RG-1}
cease to exist for $\Omega>\Omega_c$. Hence, black hole solutions are
restricted to masses $M$ with
\begin{equation}\label{Oc}
\Omega\le \Omega_c\quad{\rm and}\quad
M\ge M_c\,.
\end{equation}
For a black hole of critical mass $M_c$ we find $\Omega(M_c)=
\Omega_c$. For such a critical black hole the inner Cauchy horizon and
the outer event horizon coincide, $x_-=x_+=x_c$ with a radius of $r_c \equiv
x_c \, r_{\rm cl}(M_c)$.  Solving $f(x_c)=0$ and $f'(x_c)=0$
simultaneously leads to the critical parameter
\begin{eqnarray}\label{acr}
\Omega_{c}&=&(d-3)\,(d-2)^{-\frac{d-2}{d-3}}\\
\label{xcr}
x_{c}&=& (d-2)^{-1/(d-3)}\,.
\end{eqnarray}
We note that \eq{acr} is of order one for all $d\ge 4$.

Next, we consider the distance function \eq{D-interpol} whose index $\gamma$
interpolates between $\gamma=1$ for large $r$ and
$\gamma=\frac{1}{2}(d-1)$ for small $r$, similar to the matchings
\eq{distance} and \eq{D-interpol2}.
We find
\begin{equation}\label{G-LO}
G(r)=
\frac{G_0\, r^\alpha}{r^\alpha+\tilde\omega\,G_0\, (r_{\rm cl}+\epsilon\,r)^{\alpha+2-d}}
\end{equation}
with $r_{\rm cl}$ and $\epsilon$ from \eq{classicalBH-def} and \eq{nonepsilon}, and
\begin{eqnarray}
\label{nonalpha2}
\alpha&=&\frac{1}{2}(d-1)(d-2)\\
\label{nonomegatilde2}
\tilde{\omega}&=&\omega\, \xi^{d-2}\,\left(\s0d2-\s012\right)^{d-2}\,.
\end{eqnarray}
Consequently,
\begin{equation}\label{f-non}
f(x)=1-\frac{x^{\alpha-d+3}}{x^{\alpha}+\Omega(1+x \epsilon)^{\alpha-d+2}}
\end{equation}
and the horizon condition becomes 
\begin{equation}\label{horizoninterpol}
x^{\alpha-d+3}=x^{\alpha}+\Omega(1+x \epsilon)^{\alpha-d+2} \,.  
\end{equation}
In Fig.~\ref{nonlin7} we plot \eq{f-non} in $d=7$ for three values of
$\Omega$. The main difference in comparison with Fig.~\ref{6dlin} is
that the limit $f\to 1$ is achieved more rapidly.

Finally, we come back to a matching with general index $\gamma$,
\eq{gen-f}, with horizon condition \begin{equation} \label{nonlinhorizon}
0=1-x^{3-d}+\Omega x^{-\gamma(d-2)}\,.  \end{equation} Again, three types of
solution are found for $\gamma > \frac{d-3}{d-2}$, corresponding to
two horizons ($x_+$ and $x_-$) for $\Omega<\Omega_c$, none for
$\Omega>\Omega_c$ and a single horizon ($x_+=x_-=x_c$) for
$\Omega=\Omega_c$. Solving $f(x_c)=0$ and $f'(x_c)=0$ simultaneously
leads to the critical parameter 
\begin{eqnarray} \label{genxc}
x_c&=&\left(1-\frac{d-3}{\gamma(d-2)}\right)^{\frac{1}{d-3}}\\ 
\Omega_c&=&\frac{d-3}{\gamma(d-2)}
\left(1-\frac{d-3}{\gamma(d-2)}\right)^{\frac{\gamma(d-2)}{d-3}-1}\,.
\end{eqnarray}
It follows that condition \eq{Oc} will hold independently of the
matching used.

Next we discuss the quantitative differences between the various
distance functions.  This relates to the limit $r\to 0$, where $f(r)$ approaches
$f\to 1$, though with different rates, see Figs.~\ref{RG-f-compare}.
Effectively, the rate is parametrised through
$\gamma$. We recall that the limit $\gamma \rightarrow \infty$
switches off gravity below the horizon $x_c$. This entails, in
\eq{genxc}, that $x_c \rightarrow 1$. This is nicely seen in
Fig.~\ref{gammaOmega} where the horizons are plotted as a function of
$\Omega/\Omega_c$ for various $\gamma$ with fixed dimensionality
$d=7$. In Fig.~\ref{gammaf}, instead, we use \eq{da} with $\gamma=1$
for $r>r_{\rm cl}$ and $\gamma>1$ for $r<r_{cl}$. At $\Omega=\Omega_c$
we show $f(r)$ for various $\gamma$, and note that the limit $f \to 1$
is approached more rapidly for larger values of $\gamma$, as
expected. We conclude that $\gamma>1$ enhances the weakening of
gravity in the limit $r \to 0$.

The above findings allow first conclusions.  The RG running 
of $G(r)$ in the regime where $r\gg
r_{\rm cl}$ has little quantitative influence on the gravitational
radius $r_s$. Interestingly, the precise RG running in the deep 
short distance regime $r\ll r_{\rm cl}$ is also largely irrelevant for 
the RG improved gravitational radius. Instead, the behaviour of 
$G(r)$ and its gradient $r\,\partial_r\,G(r)$ in the regime between 
$r\approx r_{\rm cl}$ and $r\approx r_s$ is mostly responsible for 
the quantitative shift from $r_{\rm  cl}$ to $r_s$.  In consequence, 
the slight differences in the distance functions \eq{distance}, \eq{linear}, 
\eq{D-UV} and \eq{D-interpol} are attributed to a slight variation 
in the underlying RG running of $G(r)$. The RG results
from \cite{Fischer:2006fz} favour moderate values for $\gamma$, as do
regularity and minimum sensitivity considerations (see
Sec.~\ref{curvature}).  In all cases studied here, the qualitative 
behaviour of the horizon structure  remains unchanged.

%
\subsection{Critical mass}\label{CM}

\begin{figure}[t]
\includegraphics[width=.95\hsize]{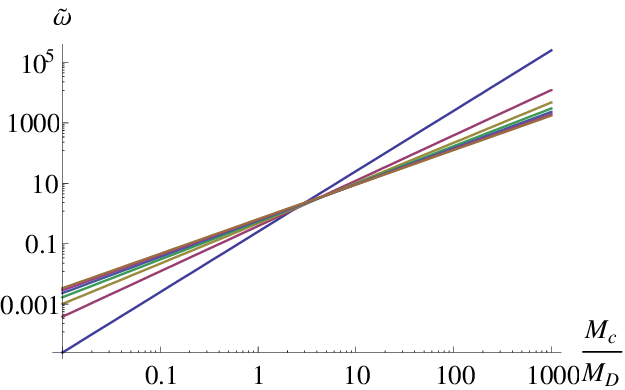}
\caption{\label{logomega} Map between the renormalisation group parameter $\tilde{\omega}$, 
the critical mass $M_c$, and the Planck mass $M_D$, based on \eq{G(k)} and
  \eq{linear} for various dimensions. From top to bottom:
  $d=4,5,\cdots,10$. }
\end{figure}

A direct consequence of our results from Sec.~\ref{horizons} is the
appearance of a lower bound on the black hole mass below which the RG
improved spacetime ceases to have a horizon, see \eq{Oc}. The critical
mass $M_c$ is defined implicitly via the simultaneous vanishing of
$f(r_s(M_c),M_c)=0$ and $f'(r_s(M_c),M_c)=0$ (here a prime denotes a
derivative with respect to the first argument). Using the solution $r_s(M)$ of
\eq{rs_implicit}, we conclude that
\begin{eqnarray}\label{Mc_implicit}
(d-3)G(r_c,M_c)&=&r_c\,G'(r_c,M_c)\,,\\
r_c&=&r_s(M_c)\,,
\end{eqnarray}
which serves as a definition for $M_c$.  The classical limit is
achieved for $M_c\to 0$. If $G(r)$ is a
monotonically increasing function of $r$, we have $r\partial_r G(r)\ge 0$.
Then, away from the classical limit, there exists a unique solution
$M_c>0$ to \eq{Mc_implicit}. Consequently, the critical mass
$M_c$  is related to the fundamental
Planck scale $M_D$ as
\begin{equation}\label{lower-1}
M_c=\zeta_c\,M_D\,.
\end{equation}
The coefficient $\zeta_c$ accounts for the renormalisation group
improvement of the black hole metric, and hence encodes the RG
effects. In the approximation \eq{G(k)}, \eq{nonlinear}, it reads
\begin{equation}
\zeta_c= \left(\frac{\tilde{\omega}}{\Omega_c}\right)^{\frac{d-3}{d-2}}
\end{equation}
where $\tilde{\omega}=\omega (\xi/c_\gamma)^{d-2}$.
The link between the RG parameters and the critical mass in units of
the fundamental Planck mass $M_c/M_D$ is displayed 
in Fig.~\ref{logomega}.
The location and the number of the horizons depends explicitly on the
value of $\Omega$, which becomes
 \begin{equation}\label{OmegaMc}
\Omega = \Omega_c \left(\frac{M_c}{M}\right)^{\frac{d-2}{d-3}}
\end{equation}
in terms of $M_c$, see \eq{sol-RG-3}. Therefore, below, we display our
results in terms of $M_c$. We return to the discussion of $M_c$ in
Sec.~\ref{DPS}.

\subsection{Horizons revisited}

Next, we return to the quantitative analysis of improved
metrics and present our numerical results for the improved
Schwarzschild radius.

\begin{figure}[t]
\includegraphics[width=.95\hsize]{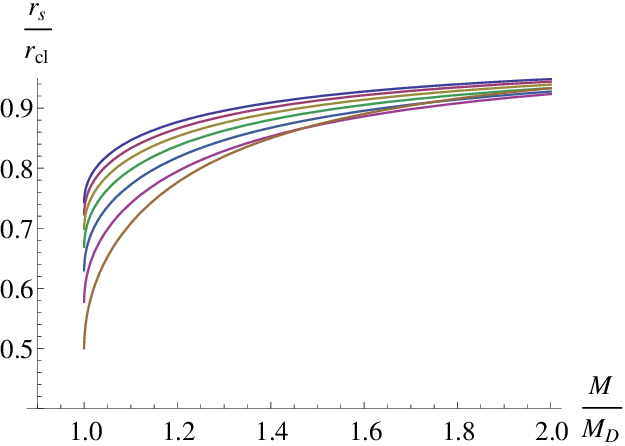} 
\caption{\label{radius}Mass dependence of the improved
  Schwarzschild radius $r_{s}(M)$ compared to its classical value
  $r_{\rm cl}(M)$; $M_c=M_D$. From bottom to top: $d=4,5,\cdots,10$. The
  end points denote the critical radii $r_c$.}
\end{figure}

\begin{figure*}
\subfigure[\label{radius2}\ Dimension and mass dependence of the horizon.]{\includegraphics[width=.45\hsize]{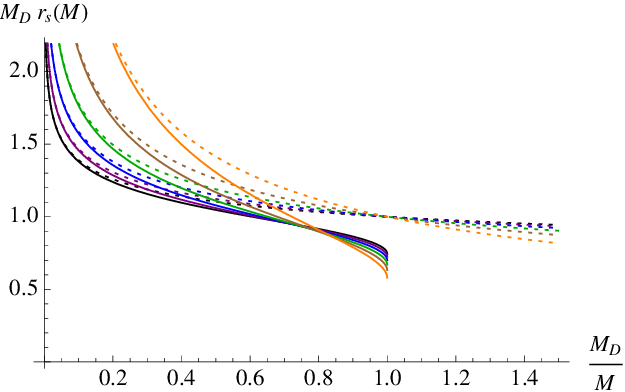}}
\hskip0.02\hsize
\subfigure[\label{logm2} \ $M_c$ dependence of Schwarzschild radius.]
{\includegraphics[width=.45\hsize]{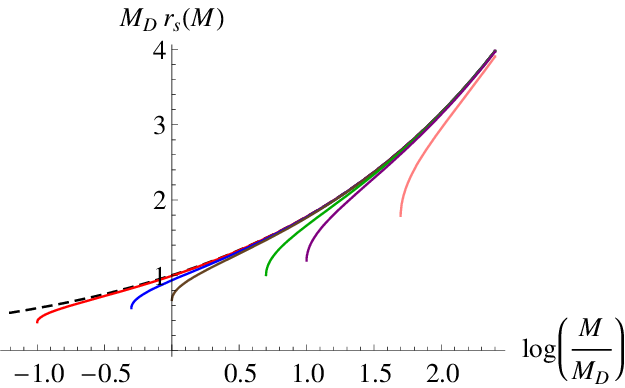}}
\caption{\label{radius-compare}Dependence of the renormalisation 
group improved Schwarzschild radius $r_{s}(M)$ 
on space-time dimension and critical mass $M_c$. 
 End points of curves
 denote the critical radius $r_c$ and dashed curves the respective classical result. 
 (a)  $M_c=M_D$ and $d=5, 6,\cdots10$, from top right to left. 
  (b) $M_c/M_D=0.1,0.5,1,5,10,50$
from left to  right.}
\end{figure*}

Figs.~\ref{radius}~\ref{radius-compare} show how the Schwarzschild radius
$r_s$ depends on the mass of the black hole $M$ in various dimensions
using \eq{linear}. In these plots we considered the scenario where the
critical mass $M_c$ is equal to the Planck mass $M_D$.  The suppression
is less pronounced with increasing dimension (Fig.~\ref{radius}). Also, the 
deviation from classical behaviour sets in at lower masses in lower dimensions, see
Fig.~\ref{radius2}.
Next we consider varying the value of $M_c$ in units of $M_D$ while keeping
the dimensionality fixed, see Fig.~\ref{logm2}. The dashed curve
corresponds to the classical result. Depending on $M_c$, quantum
corrected curves start deviating visibly as soon as the black hole
mass is only a few $M_c$ or lower. At fixed $M/M_D$, the deviation
from classical behaviour sets in earlier for larger $M_c$.

The horizon is slightly sensitive to the distance function \eq{match}, or
equivalently, to the parameter $\gamma$.  Here, $\gamma$ parametrises
how rapidly $G(r)$ weakens in the cross-over regime at scales
$r\approx r_{\rm cl}$. This can be seen from Fig ~\ref{gammaOmega}. In
the decoupling limit $\gamma \rightarrow \infty$, the critical radius
$x_c \rightarrow1$ reaches the classical value. In this
limit, gravity is switched off below $r_{\rm cl}$, implying that the
Schwarzschild radius remains unchanged. For $\gamma=1$, 
instead, the outer horizon $x_{+}$ decreases
rapidly as $\Omega$ increases towards $\Omega_c$.

In conclusion, the quantitative reduction of $r_s/r_{\rm cl}$ by
quantum effects can be associated to the behaviour of the running
coupling $G(r)$ and its decrease $r G'(r)$ at length scales set by the
horizon $r\approx r_s$. This decrease, in turn, can be understood via
the parameter $\gamma$ which controls how quickly quantum effects are
turned on as $r /r_{\rm cl}$ becomes small. For all matchings $k(r)\propto1/D(r)$
discussed in this paper, and for dimensionality $d\geq4$, the same
qualitative behaviour is observed. In particular the RG improvement
indicates that quantum black holes display a lower
bound \eq{lower-1} of the order of the Planck scale.

\subsection{Perturbation theory}\label{pt}

In the limit $M_D/M \ll 1$, quantum corrections become perturbative
and we can perform a large mass expansion. In particular, for
asymptotically heavy black holes we find $x_+ \rightarrow 1$, as can
be seen in Fig.~\ref{logm2} where $r_s$ approaches its classical value
for large $M$ and the dimensionless inner horizon $x_{-} \rightarrow
0$ in the large-mass limit. We note that the parameter $\Omega$ scales
as $\sim M^{-\frac{d-2}{d-3}}$. Hence a large mass expansion
corresponds to an expansion in $\Omega \ll 1$. In general, and
independently of the RG running and the matching, we find
\begin{equation}
\label{a} 
x_\pm=\sum_{n=0}^\infty a^\pm_n\,\Omega^n 
\end{equation} 
with dimensionless coefficients $a_n$, where $a^+_0=1$ and $a^-_0=0$. The
expansion converges rapidly, see Fig.~\ref{ab}. Explicitly, the first
few coefficients read
\begin{eqnarray} 
\label{linearx+}
x_{+}&=&1-\frac{1}{d-3} \Omega - \frac{d-2}{2(d-3)^2} \Omega^2 
 \nonumber\\ &&
 - \frac{(d-1)(d-2)}{3(d-3)^3} \Omega^3+O(\Omega^4)\quad\quad
 \\
\label{linearx-}
x_{-}&=& \Omega +\Omega^{d-2}+(d-2) \Omega^{(2d-5)} + \cdots\quad\quad
\end{eqnarray}
using the matching \eq{linear} and \eq{sol-RG-1}.  The leading order
quantum effects modify the Schwarzschild radius $r_s=x_+\, r_{\rm cl}$
and the Cauchy horizon $r_{w}=x_{-}\, r_{\rm cl}$ as
\begin{eqnarray} \label{largemhorizon}
r_s&=& r_{\rm cl}- \frac{\Omega_c}{d-3}
\left(\frac{M_c}{M_D}\right)^{\frac{d-2}{d-3}} \frac{1}{M} + {\rm
  subleading} \,,\quad\quad\quad \\ r_{w}&=&\Omega_c
\left(\frac{M_c}{M_D}\right)^{\frac{d-2}{d-3}} \frac{1}{M} + {\rm
  subleading.}
\end{eqnarray} 
Thus, in the limit $M_D/M \rightarrow 0$ we confirm $r_s \rightarrow
r_{\rm cl}$ and $r_w \rightarrow 0$, as expected.

Interestingly the inner horizon behaves differently if we employ the
non-linear matching \eq{D-interpol}. To that end, we again solve the
horizon condition, now given by \eq{horizoninterpol},  and expand in
$\Omega \ll 1$ to find $x_+$ and $x_-$ to leading order in $\Omega$,
\begin{eqnarray}
x_{+}&=&1- \frac{(1+\epsilon)^{\alpha-d+2}}{d-3} \Omega + {\rm
  subleading}\quad\quad \\ x_{-}&=& \Omega^{\frac{1}{3-d+\alpha}}+
{\rm subleading}
\end{eqnarray}
where $\alpha$ and $\epsilon$ are given by \eq{nonalpha2} and
\eq{nonepsilon}. We note that if we take $\alpha= d-2$ we recover
\eq{linearx+} and \eq{linearx-}. In the non-linear case the outer
horizon $r_s$ has a large mass expansion similar to
\eq{largemhorizon}, whereas the inner horizon has a large mass
expansion whose leading term is proportional to a positive
power of the mass,
\begin{eqnarray}
r_s&=& r_{\rm cl}- \frac{\Omega_c (1+\epsilon)^{\alpha-d+2}}{d-3}
\left(\frac{M_c}{M_D}\right)^{\frac{d-2}{d-3}}
\frac{1}{M}\quad\quad\quad \\ r_{w}&=&\0{1}{M_D} \left(\frac{M_c
  \Omega_c}{M_D}\right)^{\rho_0} \left(\frac{M}{M_D}\right)^{\rho_1}
\end{eqnarray} 
plus terms subleading in $M$. We have introduced $\rho_0=
\frac{d-2}{(d-3)(\alpha +3 -d)}$,
$\rho_1=\frac{5-2d+\alpha}{(d-3)(\alpha+3-d)}$ and
$\rho_1+\rho_2=\01{d-3}$.  For $d=4$, $\rho_{1}=0$ for $d>4$ we find
$1>\rho_1>0$. This implies that in the limit $M \rightarrow \infty$
the Cauchy horizon $r_w$ will approach a constant for $d=4$. In higher
dimensions $d>4$, $r_w$ will increase with mass as $r_{w} \sim
M^{\rho_1}$, whereas the ratio $r_{w}/r_{s} \sim M^{2-d} \,
\rightarrow 0$ in the large mass limit.

\subsection{Threshold effects}\label{te}

\begin{figure*}
\subfigure[\ Perturbative approximation of the horizon.]
{\includegraphics[width=.45\hsize]{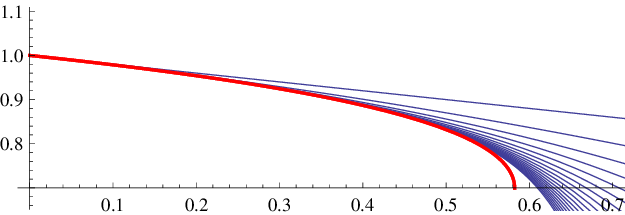}}
\hskip0.09\hsize
\subfigure[\ Threshold approximation of the horizon.]
{\includegraphics[width=.45\hsize]{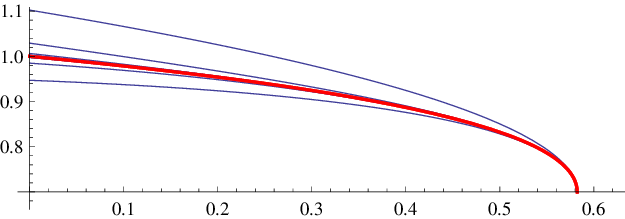}}
\caption{\label{ab} Location of the event horizon $x_+(\Omega)\equiv r_s/r_{\rm cl}$ 
as a function of the parameter $\Omega$ in $d=8$ (thick line)
  within  various approximations (thin lines). (a)
  Perturbative expansion about $\Omega=0$ using \eq{a} at order $n=1,2,\cdots,20$, 
  approaching the exact solution adiabatically (from top to bottom). 
  (b) Threshold expansion about the critical point $\Omega=\Omega_c$ using \eq{b} at order $n=1,2,\cdots,9$, 
 alternating towards the full solution.}
\end{figure*}

The RG improved black hole displays an interesting threshold behaviour
in the vicinity of $M\to M_c$. This can be understood as
follows. Suppose we read \eq{f-RG} as a function of $r$ and $M$,
$f(r,M)$, and perform a Taylor expansion in both variables. The
outcome is then evaluated at the horizon $r=r_s(M)$ where
$f(r_s(M),M)=0$. Independently of the chosen expansion point
$(r_0,M_0)$ with $r_0=r_s(M_0)$, we find 
\begin{eqnarray}\nonumber 
0&=&
\sum_{n=1}\left[ \frac{1}{n!}(M-M_0) ^n\frac{\partial^n f}{\partial
    M^n}
\label{threshold}
+\frac{1}{n!}(r-r_0)^n\frac{\partial^n f}{\partial r^n}\right] 
\end{eqnarray} 
at the horizon. Note that the derivatives are evaluated at
$(r,M)=(r_0,M_0)$. If the RG running of $G$ is $M$-independent, the
expansion has only a linear term in $(M-M_0)$. At threshold where
$r_0=r_c$, we furthermore have $\partial f/\partial r|_{r_c} = 0$. In
addition, $\partial f/\partial M|_0$ is always non-zero. Therefore,
close to $M\approx M_c$, $f(r(M),M)=0$ can only be satisfied if
\begin{equation}\label{sqrt} 
r_s(M)-r_c\sim \sqrt{M-M_c}\,, 
\end{equation} 
provided that $\partial^2 f/\partial r^2|_{r_c} \neq 0$.  More generally, if the
first non-vanishing derivative $\partial^n f/\partial r^n|_{r_c}$
occurs at order $n$, the threshold behaviour \eq{sqrt} becomes
\begin{equation}
\label{higher} r_s(M)-r_c\sim \sqrt[n]{M-M_c}\,.  
\end{equation} 
The generic case encountered in this paper, for all RG runnings employed, is
$n=2$.  Consequently, at threshold, we have the expansions
\begin{equation}
\label{b} 
x_\pm=\sum_{n=1}^\infty b^\pm_n\,
\left(\frac{M}{M_c}-1\right)^{n/2} 
\end{equation} 
with dimensionless coefficients $b^\pm_n$. {This is equivalent to an
expansion in powers of $\sqrt{\Omega_c-\Omega}$.} This expansion
converges rapidly as can be seen from Fig.~\ref{ab}, where the first
few terms (up to $n=6$) are enough to match the full solution even for
small $\Omega$.

Explicitly, using the matching \eq{linear}, the behaviour \eq{sqrt}
reads to leading order
\begin{equation}
r_s(M)-r_c = 
\frac{(G_0 M_c)^{1/(d-3)}}{ \sqrt{\012(d-3)M_c}} \sqrt{M-M_c}\,.
\end{equation}
In the light of the above, the dependence of the horizon radius on the
black hole mass can easily be understood, see Fig.~\ref{radius}. For
large black hole mass $M\gg M_c$, $\partial_r f$ is non-vanishing at
$f(r_s)=0$, implying that the linear terms in \eq{threshold} have to
cancel. This leads to the very soft dependence of $r_s/r_{\rm cl}$ on
$M_c/M$ for large $M$. With decreasing $M$, $\partial_r f$ is
decreasing as well, thereby increasing the admixture from $(r-r_0)^2$
corrections. The latter fully take over in the limit $M\to M_c$,
leading to non-analytical behaviour \eq{sqrt} which is nicely seen in
Fig.~\ref{radius}. We stress that this structure is independent of the
dimension as long as $d>3$.

\subsection{Temperature and specific heat}\label{t}
The threshold behaviour of Sec.~\ref{te} has important implications
for the thermodynamics of black holes. Based on the RG improvement
used here, the Bekenstein-Hawking temperature is given by \cite{Hawking:1979,Gibbons:1976ue} 
\begin{equation}
T=\frac{f'(r_s(M),M)}{4\pi}\,,
\end{equation}
where the prime denotes a derivative with respect to the first
argument. We conclude that the temperature is bounded, $T(M)\le T_{\rm cl}(M)$. Equality is
achieved for asymptotically large black hole mass, where the
temperature scales inversely with mass, $T\sim 1/M$, and the specific
heat
\begin{equation}
C=\frac{\partial M}{\partial T}
\end{equation} 
is negative. In this regime, it is expected that the black hole evaporates 
through the emission of Hawking radiation, thereby lowering its mass 
through the influx of negative energy.
At threshold with $M\to  M_c$, however, the black hole
temperature vanishes because $f'(r_c)\to 0$. In the vicinity of
$M\approx M_c$, the temperature of the black hole scales
non-analytically, 
\begin{equation} 
\frac{\partial T}{\partial M}\sim
\frac{\partial r_s(M)}{\partial M}\sim \frac{1}{\sqrt{M-M_c}}>0 
\end{equation}
as a consequence of \eq{sqrt}. Therefore, the specific heat has become
positive, and the temperature displays a maximum $T(M)\le T_{\rm max}$
for all $M$. Furthermore, since $\partial T/\partial M$ vanishes at
$T_{\rm max}$, the specific heat changes its sign through a pole.  We
conjecture that black holes with mass $M_c$ constitute cold remnants,
provided they can be reached through an evaporation process. This would
require, however,  that $T_{\rm max}$ remains below the fundamental Planck scale.

\subsection{Renormalisation and the Planck scale}\label{DPS}

We  summarise our results.
The main physics of this paper originates from a 
new mass scale $M_c$, which is absent in the classical theory.
Its existence is due to
quantum gravity corrections, implemented on the level of the
metric. 

For black hole mass $M$ large compared to  $M_c$, renormalisation group 
corrections to the metric are small. The
gravitational force remains strong enough to allow for black
holes. Improved black hole metrics display horizons of the
order of the classical horizon, the specific heat stays negative and the temperature 
scale inversely proportional to mass, modulo small quantum corrections. This is the
semi-classical regime of the theory.

Once  the mass $M$ approaches $M_c$, we observe the transition from 
strong to weak gravity. In its vicinity, renormalisation effects become of 
order one, the reduction of the event horizon becomes more pronounced, 
the specific heat has become positive and the
thermodynamical properties are no longer semi-classical. This is the 
Planckian (or quantum) regime of the theory.

When $M$ drops below $M_c$, renormalisation 
group corrections to the metric have become strong. The gravitational force 
has weakened significantly, to the point that improved black hole space-times 
no longer display a horizon. This is the deep UV scaling regime of the
theory. The improved metric differs qualitatively from the classical one. Hence, the 
applicability of our renormalisation group improvement 
becomes doubtful, and conclusions from this regime
have to be taken with care.

 If the renormalisation
effects of the black hole space-time are parametrically
strong, $\tilde\omega\gg 1$,
the scale $M_c$ grows large, and parametrically larger than the
Planck scale $M_D$. In turn, for weak renormalisation
$\tilde\omega\ll 1$, the scale $M_c$ remains small as well. 
We note that $M_c$ vanishes in the classical limit where
quantum corrections are switched off. The reason for this is 
that the Schwarzschild
solution of classical general relativity does not predict its own
limit of validity under quantum corrections. Interestingly, the underlying fixed point is 
not primarily responsible for the existence of the lower bound $M_c$. Other
ultraviolet completions of gravity such as string theory, loop quantum gravity or 
non-commutative geometry can lead a similar weakening of the gravitational force at 
length scales of the order of the Planck length.

To conclude, the improved metric changes  qualitatively at $M\approx
M_c$.  Therefore it is tempting to interpret $M_c$ as a `renormalised' Planck
scale. Its numerical value  depends on the precise renormalisation group running.
As long as the latter is driven by the 
gravitational self-coupling only, it is natural to have $M_c$  of
the order of $M_D$. This may be different once strong renormalisation effects are 
induced by external mechanisms, eg.~through the coupling to a large number 
of matter fields.

\section{Space-time structure and Penrose diagram}\label{STS}

In this section, we study the implications of quantum gravitational
effects on the space-time structure of black holes, including a
discussion of critical black holes, an analogy with
Reissner-Nordstr\"om black holes, an interpretation in terms of an
effective energy momentum tensor, the (absence of) curvature
singularities at the origin, and the causality structure and Penrose
diagram of quantum black holes.

\subsection{Critical black holes}

The space-time structure of a critical black hole with mass $M=M_c$
has a single horizon at $r_c=r_{\rm cl}\,x_c$ and $x_{-}=x_{+}=x_c$
where the function $f(x)$ has a double zero $f(x_c)=0$. For the
matching \eq{linear} $x_c$ is given explicitly by \eq{xcr}. The
near-horizon geometry of a critical black hole is obtained by
expanding $f(x)$ around $x=x_c$. We find
\begin{equation}
f(x)=  \0{1}{2} \, \bar{x}^2\, f''(x_c,\Omega_c)
\end{equation} 
where $\bar{x}\equiv x-x_c$ and the double prime represents a second
derivative with respect to $x$. Therefore, we can write the line
element in terms of the coordinate $\bar{r}=r-r_c$ as
\begin{eqnarray}
\label{AdS}
ds^2&=& 
-\frac{\bar{r}^2}{r_{\rm AdS}^2}dt^2
+ \frac{r_{\rm AdS}^2}{\bar{r}^2} dr^2 
+ r_c^2\, d\Omega^2_{d-2}\quad
\end{eqnarray}
The metric \eq{AdS} is the product of a two-dimensional anti-de Sitter
space with a $(d-2)$-sphere, ${\rm AdS}_2 \times S^{d-2}$, and depends
on their respective radii
\begin{eqnarray}
\label{rAdS}
r_{\rm AdS}&=&(G_0 M_{\rm AdS})^{{1}/{(d-3)}}\\
r_c&=& x_c\,(G_0\,M_c)^{1/(d-3)}
\end{eqnarray}
The curvature of the anti-de Sitter part is determined by the mass
parameter $M_{\rm AdS}$,
\begin{equation}\label{AdSmass}
M_{\rm AdS}=  M_c  \left(\frac{1}{2}f''(x_c, \Omega_c)\right)^{-\012(d-3)}\,.
\end{equation}
Using \eq{linear} for $d=4$ we have $M_{\rm AdS}=
\frac{1}{\sqrt{2}}M_c$. For higher dimensions $d=6,8, 10$ we find
$M_{\rm AdS}/M_c\approx 0.14, 0.017, 0.0016$, respectively. For all
dimensions, we have $M_{\rm AdS}<M_c$. We note that the metric
\eq{AdS} is of the form of a Robinson-Bertotti metric for a constant
electric field.

\subsection{Reissner-Nordstr\"om-type metrics}

It is interesting to compare the RG improved black hole with the
well-known Reissner-Nordstr\"om solution of a charged black
hole in higher dimensions \cite{Myers:1986un}. The latter is defined
via the lapse function
\begin{equation}\label{fRN}
f_{\rm RN}(r)=1-\frac{G_0 M}{r^{d-3}} + \frac{G_0 e^2}{r^{2(d-3)}}
\end{equation}
where $e^2$ denotes the charge of the black hole (squared). The charge
has the mass dimension $[e^2]=4-d$.  The physics of the
Reissner-Nordstr\"om black hole is best understood in terms of the
dimensionless parameter
\begin{equation}\label{oRN}
\Omega_{\rm RN}=\frac{e^2}{G_0 M^2}
\end{equation}
which measures the relative strength of the competing terms on the rhs
of \eq{fRN}. In terms of \eq{oRN} and using $x=r/r_{\rm cl}$ , the
lapse function becomes
\begin{equation}\label{RN}
f_{\rm RN}(x)=1-\frac{1}{x^{d-3}}+\frac{\Omega_{\rm RN}}{x^{2d-6}}\,.
\end{equation}
For $\Omega_{\rm RN}>\frac{1}{4}$ the spacetime has no horizons and
exhibits a naked singularity. For $\Omega_{\rm RN}<\frac{1}{4}$ the
spacetime displays two horizons, whereas for $\Omega_{\rm RN} =
\frac{1}{4}$ the black hole displays a single horizon. Therefore
$\Omega_{\rm RN}=\frac{1}{4}$ is referred to as a extremal black hole
with critical mass $M_{\rm RN, c}=2\sqrt{e^2/G_0}$. The radius of the
extremal black hole is given by $r_{\rm RN,c}= 2^{-1/(d-3)}\, r_{\rm
  cl}$.

Reissner-Nordstr\"om spacetimes share some of the qualitative features
of RG improved higher dimensional black holes discussed in this paper.
If we consider a quantum black hole using the matching \eq{linear} and
expand the lapse function to leading order in $\Omega$, we find
\begin{equation}\label{LO}
f_{\rm LO}(x)=1-\frac{1}{x^{d-3}}+\frac{\Omega}{x^{2d-5}}+{\rm subleading}\,.
\end{equation}
In either case \eq{RN} and \eq{LO}, the relevant physics originates
from competing effects: a leading order Schwarzschild term
$~\,-1/r^{d-3}$, which is counterbalanced by either the charge,
parametrised by $\Omega_{RN}\sim e^2$, or by quantum corrections due
to a running gravitational coupling, parametrised by
$\Omega\sim\tilde{\omega}$. The correction terms become quantitatively
dominant with decreasing $r \rightarrow 0$. We note that \eq{RN} and
\eq{LO} are formally equal for $\Omega_{\rm RN}=\Omega \, x$. In
either case, in the large mass limit $M \rightarrow \infty$ we find an
outer horizon $f(x_{+})=0$ for $x \approx 1$. It follows that the near
horizon geometry of a quantum black hole is approximately that of a
Reissner-Nordstr\"om black hole of charge $e^2=\tilde{\omega}\,
r_{\rm cl}^{d-4}$, and in the large mass limit.

Next we consider the near horizon geometry of an extremal
Reissner-Nordstr\"om black hole, which is of the ${\rm AdS}_2 \times
S^{d-2}$ type. The line element is given by \eq{AdS} where
\begin{eqnarray}
 \label{AdSradiusRN}
r_c&=&\left(\s012 G_0 M_{\rm RN, c}\right)^{1/(d-3)}\\
\label{AdSmassRN}
M_{\rm AdS}&=&  M_{RN, c}\,  \left(\s0{1}{2}f_{RN}''(x_c, \Omega_c)\right)^{{-(d-3)}/{2}}\,.\quad
\end{eqnarray}
 For $d=4$ we find $M_{\rm AdS}= \frac{1}{2}M_{\rm RN, c}$. For higher
 dimensions $d=6, 8$ and $10$, we obtain $M_{\rm AdS}/M_{\rm RN, c}
 \approx 0.02, 0.0002$ and $ 6 \times 10^{-7}$, respectively. The
 decreasing of $M_{\rm AdS}/M_{\rm RN, c} $ with dimension is similar
 to the decreasing of $M_{\rm AdS}/M_{c}$ for the critical black hole
 \eq{AdSmass}.

\subsection{Effective energy-momentum tensor}

In this paper we have obtained our results by replacing the classical
value of Newton's constant $G_0$ with a running constant $G(r)$.  It
is interesting to ask whether this modification could have arisen from
an explicit source term, the energy-momentum tensor, for Einstein's
equations. The answer is affirmative, and obtained by inserting the RG
improved metric into the left hand side of the Einstein equations
$G^{\mu \nu}= 8 \pi\, G_0\, T^{\mu \nu}$. The non-vanishing components
are the diagonal ones $T^\mu_{\, \nu}= {\rm diag}(-\rho, p_r,
p_\perp,..,p_\perp)$, given by 
\begin{equation} \rho=-p_r=\frac{G'(r)\,M_{\rm
    phys}}{S_{d-2}\, G_0\, r^{d-2}} 
    \end{equation} 
    \begin{equation}
p_\perp=-\frac{G''(r)\,M_{\rm phys}}{(d-2) S_{d-2} G_0 r^{d-3}} 
\end{equation}
where $S_{d-2}={2 \pi^{(d-1)/2}}/{\Gamma((d-1)/2)}$.  Integrating the
energy density $\rho$ over a volume of radius $r$ one finds the
effective energy within that radius, 
\begin{equation} E(r)=S_{d-2} \int^r_0 dr'
\rho(r')r'^{d-2}= \frac{G(r) \,M_{\rm phys}}{G_0} \, , 
\end{equation} 
where we assume $G(r)$ obeys the limits $G(r)\rightarrow 0$ for
$r\rightarrow0$ and $G\rightarrow G_0$ for $r\rightarrow \infty$.  As
such we note that $E(\infty)=M_{\rm phys}$ the physical mass.  We also
make the observation that replacing $G(r)\,M_{\rm phys} \rightarrow
G_{0}E(r)$ leaves the metric invariant.

\subsection{Absence of curvature singularities} \label{curvature}
In this section, we discuss the $r\to 0$ limit of asymptotically safe 
black holes and the absence of curvature singularities. Classical
Schwarzschild solutions display a coordinate singularity at $r=r_{\rm
  cl}$ where $f(r)^{-1} \rightarrow \infty$. Curvature invariants
remain well-defined and finite at the horizon, which shows that the
singularity is only an apparent one.

A curvature singularity in the classical metric is found at $r\to 0$,
where $f(r)\rightarrow -\infty$ and the Ricci scalar diverges as $R
\sim r^{1-d}$. This curvature singularity implies the break-down of
classical physics at the centre of a black hole. It is expected that
quantum fluctuations should lead to a less singular or finite
behaviour as $r \rightarrow 0$. 

Within the renormalisation group
set-up studied here, the main new input is the anti-screening of the
gravitational coupling.
Consequently, $G(r)/G_N$ becomes very small, thereby modifying the
$r\to 0$ limit. For small $r/r_{\rm cl}\ll 1$, we have
\begin{equation}\label{smallr}
f(r) = 1-(\mu\,r)^\sigma\,+\,{\rm subleading}
\end{equation}
where the mass scale $\mu$ and the parameter $\sigma$ are fixed by the
renormalisation group and the matching condition discussed in
Sec.~\ref{QG}. We note that a value of $\sigma=2$ would correspond to
a de Sitter core with cosmological constant
\begin{equation}\label{LdS}
\Lambda_{\rm dS}=\frac{1}{2}(d-1)(d-2)\mu^2\,.
\end{equation}
More generally, the value of $\sigma$ depends on the detailed short
distance behaviour. We find $\sigma_{\rm ir}=1$ using \eq{linear} for
all values of $d\geq4$, and $\sigma_{\rm uv}=\s012(d^2-5d+8)$ using
\eq{D-UV}.  For $d\geq4$ the latter takes values $\sigma_{\rm
  uv}\geq2$. In contrast to this, the classical solution displays
$\sigma_{\rm cl}= 3-d$. Using the matching \eq{nonlinear} with
parameter $\gamma$ we have $\sigma=\gamma(d-2)-d +3$.  Consequently, a
de Sitter core is achieved for
\begin{equation}
\gamma_{\rm dS}=(d-1)/(d-2)\end{equation} 
in the limit $r \rightarrow 0$.

Next, we calculate the Ricci scalar, the Riemann tensor squared and
the Weyl tensor squared in the limit $r\rightarrow0$, using
\eq{smallr}. The results are
\begin{eqnarray}\nonumber
R&=& F_R\cdot(\mu\, r)^{\sigma-2}\,\mu^2\\
R^{\mu\nu\kappa\lambda}R_{\mu\nu\kappa\lambda}&=&F_{\rm Riem}\cdot(\mu\,
r)^{2\sigma-4}\,\mu^{4}\nonumber \\
C^{\mu\nu\kappa\lambda}C_{\mu\nu\kappa\lambda}&=&F_C\cdot(\mu\,
r)^{2\sigma-4}\,\mu^{4}
\end{eqnarray}
modulo subleading corrections. The coefficients are
\begin{eqnarray}
F_R&=& (\sigma+d-2)(\sigma+d-3)\nonumber \\
F_{\rm Riem}&=&\sigma^4-2\sigma^3+(2d-3)\sigma^2+2(d-2)(d-3)\nonumber\\
F_C&=&\frac{d-3}{d-1}(\sigma-1)^2(\sigma-2)^2
\end{eqnarray}
Clearly, the curvature singularity is absent as soon as $\sigma\ge 2$,
which in general is achieved for the matchings employed here
including \eq{D-UV}.  For the matching \eq{linear}, however, we have
$\sigma=1$ and conclude that in this case the remaining curvature
singularity reads $R \sim \frac{1}{r}$. This is still a significant
reduction in comparison with the behaviour $\sim r^{1-d}$ within the
classical Schwarzschild solution, and indicates that the weakening of
gravitational interactions leads to a better short distance behaviour.

The Riemann-squared coefficient is non-zero for all values of $\sigma$
when $d\geq4$. The Weyl-squared term has the same $r$-dependence as
the Riemann -squared term, but its coefficient vanishes for both
$\sigma=1$ and $\sigma=2$. Hence, there is no choice for $\sigma$
which makes all three coefficients vanishing.

We are lead to the following conclusions. Regularity of an asymptotically safe 
black hole requires $\sigma\geq 2$. The RG study indicates that the behaviour 
for the physical theory lies in between the limits set by $\sigma_{\rm ir}
\le \sigma_{\rm  phys}\le\sigma_{\rm uv}$. It is tempting to speculate that the
physical value would read $\sigma=2$ corresponding to a de Sitter core
with positive cosmological constant set by \eq{LdS}. A distance function 
with effective index $\gamma\ge\gamma_{\rm dS}$ together with 
a momentum-scale RG for Newton's coupling provides for a 
singularity-free metric for all $r$. This is a very mild constraint on the RG 
running, as $\gamma_{\rm dS}\in[1,\frac{3}{2}]$ is very close 
to $\gamma_{\rm ir}=1$ for all $d\ge4$.

As a last observation, we treat $\gamma$ as a free parameter and
employ a principle of minimum sensitivity (PMS) condition to identify
its best match value $\gamma_{\rm PMS}$ \cite{Stevenson:1981vj}. Since
the horizon radius $r_s=r_s(\gamma)$ at fixed black hole mass depends
monotonically on $\gamma$, a PMS condition singles out the boundaries
of the parameter space given by the decoupling limit $1/\gamma\to 0$,
and the de Sitter limit $\gamma=\gamma_{\rm dS}$. The decoupling limit
corresponds to the switching-off of gravity. Hence, we conclude that a
minimum sensitivity condition singles out $\gamma_{\rm
  PMS}=\gamma_{\rm dS}$.

\subsection{Kruskal-Szekeres coordinates}

In this section we introduce Kruskal-Szekeres coordinates which remove
the coordinate singularities at the horizons. This is the first step
towards a discussion of the causal structure of asymptotically safe black holes
and their Penrose diagrams. 

Here we consider the case where $M>M_c$
such that the spacetime has two horizons; the outer horizon $r_s
\equiv r_{\rm cl}\,x_+$ and the Cauchy horizon $r_w \equiv r_{\rm
  cl}\,x_{-}$. For simplicity we will consider the linear matching
\eq{linear} where the lapse function is given by \eq{f-L} such that
$\alpha=d-2$. The horizons are found by the real positive roots of
\eq{sol-RG-1}. In general there will be exactly $\alpha=d-2$, possibly
complex, roots. In the regime of interest where $0< \Omega<\Omega_c$,
we have always two real positive roots $x_\pm$. In even or odd
dimensions, we additionally find $(d-4)/2$ pairs of complex conjugate
roots, or a real negative root and $(d-5)/2$ pairs of complex
conjugate roots, respectively. Therefore, we decompose
\begin{eqnarray}
\Delta &\equiv& x^{\alpha}+\Omega-x \nonumber\\
&=& (x-x_{+})(x-x_-)\prod_{i=1}^{\alpha-2}(x-z_i).
\end{eqnarray}
into the two real roots $x_{\pm}>0$ and the remaining $d-4$ roots
$z_i$. In terms of these, we have
\begin{equation}
\Omega= (-1)^{\alpha}\, x_+\, x_-\, \prod_{i=1}^{\alpha-2} z_i \,.
\end{equation}
We express the line element in terms of the roots and the
dimensionless radial coordinate $x$
\begin{equation}
ds^2=-\frac{\Delta}{x^\alpha+\Omega}dt^2 
+ \frac{x^\alpha+\Omega}{\Delta} dr^2 +  r^2\ d\Omega^2_{d-2} \, .
\end{equation}
Next we express the line element in terms of Kruskal-Szekeres type
coordinates to remove the coordinate singularities. We will follow the
method as outlined in \cite{Townsend:1997ku} for a Reissner-Nordstr\"om black
hole with two horizons. First we define the dimensionless tortoise
coordinate
\begin{equation}
dx^{*}=\frac{x^\alpha+\Omega}{\Delta}\,dx
\end{equation}
It is then clear that radial null geodesics correspond to $t/r_{\rm
  cl} = \pm x^*$.  Performing the integral we find
\begin{eqnarray}
x^*&=&x+\frac{1}{2\kappa_+} \ln( \left|x-x_+\right|) + \frac{1}{2\kappa_-} \ln(\left|x-x_-\right|) 
\nonumber\\ &&
\label{xstar}
+ \sum^{\alpha-2}_{i=1} \frac{1}{2\kappa_i} \ln((x-z_i))+ {\rm constant}\\
\kappa_i&=&\frac{(z_i-x_+)(z_i-x_-)\prod^{\alpha-2}_{j \neq i} (z_i-z_j)}{2z_i}\quad \ \ \ \\
\kappa_+&=&\frac{(x_+-x_-)\prod^{\alpha-2}_{j=1} (x_+-z_j)}{2x_+}\\
\kappa_-&=&\frac{(x_{-}-x_+)\prod^{\alpha-2}_{j=1} (x_--z_j)}{2x_-}
\end{eqnarray}
We now introduce advanced and retarded time coordinates given by
\begin{eqnarray}
v=x^*+w\\
u=x^*-w
\end{eqnarray}
where $w$ is the dimensionless time $w \equiv t/r_{\rm cl}$. We then
define the coordinates
\begin{eqnarray}\label{KS}
V^{\pm}=e^{\kappa_\pm\,v}\\
U^{\pm}=-e^{\kappa_\pm\,u}
\end{eqnarray}
These are the KS-type coordinates for quantum black holes. The product
\begin{equation} U^\pm V^\pm= -e^{2\kappa_\pm x^*} \end{equation} is a constant for any
given radius $x$.  In terms of the coordinates $U^+$ and $V^+$ the
line element becomes \begin{eqnarray} ds^2=&-&\left(\frac{r_{\rm
    cl}}{\kappa_{+}}\right)^2 \, e^{-2 \kappa_{+}\,x^*}
\frac{\Delta}{x^\alpha+\Omega} dU^+dV^+ \quad \ \ \ \nonumber\\ &+&
r^2\ d\Omega^2_{d-2} \end{eqnarray} Inserting $x^*$ given by \eq{xstar} we find
\begin{eqnarray}
ds^2&=&-r_{\rm cl}^2 \, F_+(x)\, dU^+\, dV^+ + r^2\ d\Omega^2_{d-2}
\\
F_{+}&=& \frac{e^{-\kappa_+x}}{x^\alpha+\Omega}\frac{\kappa_+^{-2}(x-x_-)}{(x-x_-)^{\frac{\kappa_+}{\kappa_-}}} \prod^{\alpha-2}_{i=1} \frac{x-z_i}{(x-z_i)^{\frac{\kappa_+}{\kappa_i}}}\,.\quad\quad
\end{eqnarray}
The singularity in the $x_+$ coordinate has been removed and the
metric covers regions of space time for $x>x_-$. There remains a
singularity at $x=x_-$, and, therefore, the metric does not cover the
region $x\leq x_{-}$. Instead we use the coordinates $U^-$ and $V^-$
in terms of which the line element is given by
\begin{eqnarray}
ds^2&=&-r_{\rm cl}^2 \, F_-(x)\, dU^-\,dV^- + r^2\ d\Omega^2_{d-2}\\
F_{-}&=& \frac{e^{-\kappa_+x}}{x^\alpha+\Omega}
\frac{\kappa_-^{-2}(x-x_+)}{(x_+-x)^{\frac{\kappa_-}{\kappa_+}}} 
\prod^{\alpha-2}_{i=1} \frac{x-z_i}{(x-z_i)^{\frac{\kappa_-}{\kappa_i}}}\,.\quad\quad
\end{eqnarray}
Hence the singularity at $x=x_{-}$ is removed in these coordinates and
the metric is well defined in the region $x<x_{+}$.  The singularity
at $x=x_{+}$ remains in this parametrisation and does not cover the
region $x>x_+$. The coordinates \eq{KS} are defined such that for
ingoing null rays $V^\pm={\rm constant}$ and for outgoing null rays
$U^\pm={\rm constant}$.

 \begin{figure}[t]
		\includegraphics[width=1.25 \hsize]{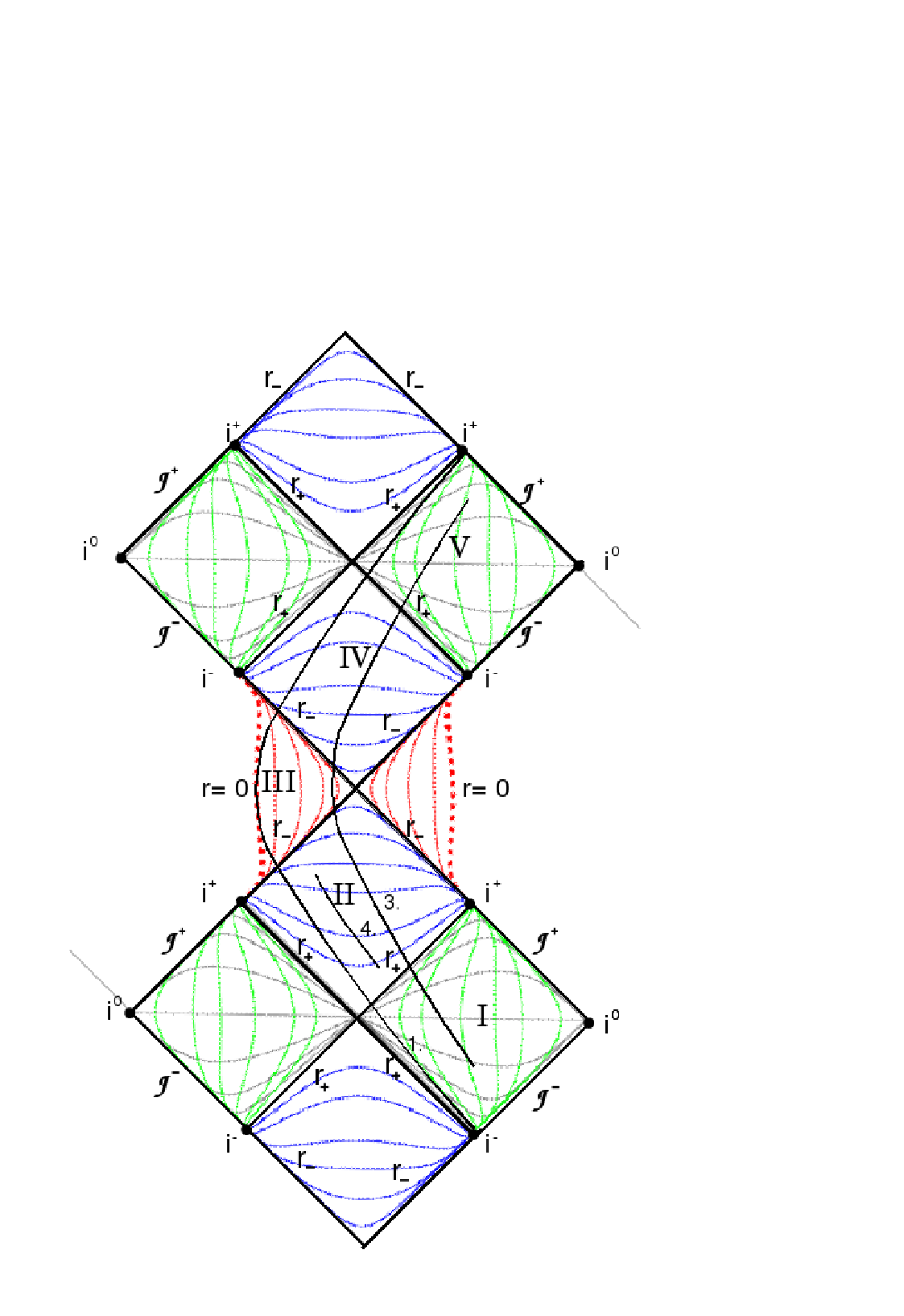}
		\caption{\label{Penrose} The Penrose diagram of a
                  quantum black hole with $M>M_c$. The black curves in
                  regions I and V are curves of constant $t$. The
                  green (blue) [red] curves are curves of constant $r$
                  in region I and V (II and IV) [III],
                  respectively. In- and outgoing radial null geodesics
                  are at $45^o$. Curves 1., 3. and 4. correspond to
                  schematic plots of various solutions to the
                  equations of motion. The points $i^0$, $i^+$ and
                  $i^-$ denote spatial infinity, future infinity and
                  past infinity, respectively.  $\mathcal{J^-}$ and
                  $\mathcal{J^+}$ denote past and future null infinity
                  (see text).}
\end{figure}

\subsection{Causality and Penrose diagram}
The global structure of the black hole can be represented by a Penrose
diagram. To produce the diagram we make an analytical continuation of
the KS-type coordinates and then map them to a finite interval
\begin{eqnarray}
V^\pm \rightarrow \tanh(V^\pm) \\
U^\pm \rightarrow  \tanh(U^\pm).
\end{eqnarray}
The resulting Penrose diagram is shown in Fig.~\ref{Penrose}. The
causal structure can be understood by noting that null geodesics are
always at $45^o$ such that ingoing photons point ``north-west'' and
outgoing photons point ``north-east''. Regions I, II and III
correspond to $x>x_+$, $x_-<x<x_+$ and $x<x_-$, respectively, where
$x_+$ denotes the outer horizon and $x_-$ the inner Cauchy horizon in
units of $r_{\rm cl}$. The other regions are the analytical
continuations; in particular regions IV and V correspond to
$x_-<x<x_+$ and $x>x_+$.  Surfaces of constant $r$ in region II
($x_-<x<x_+$) are trapped surfaces such that all null geodesics move
towards the inner horizon.  On the other hand region IV defines a
white hole where all null geodesics point towards $r_+$.

To get an idea of the causal structure experienced by an in-falling
observer we follow the standard procedure of considering a radially
moving test particle as was done in the $d=4$ case
\cite{Bonanno:2000ep}. We define the dimensionless proper time of the
radial particle $d\tau^2=ds^2/r_{\rm cl}^2$. A constant of motion
$\zeta$ is defined by the Killing vector equation corresponding to the
time independent nature of the metric,
\begin{equation}
\zeta = f(x) \frac{dw}{d\tau} \, .
\end{equation}
From the form of the metric \eq{ds2} the equations of motion for the test particle can then be given 
in terms of $\zeta$:
\begin{equation}\label{eqmotion}
\dot{x}^2= \zeta^2-f(x)
\end{equation}
(dots denote derivatives with respect to proper time $\tau$.)  We
define a Newtonian-like potential, $\Phi(x)=\frac{1}{2}(f(x)-1)$, to
write an equation for the proper acceleration of the test particle
\begin{equation}
\ddot{x}=-\frac{\partial \Phi(x)}{\partial x}\, .
\end{equation}
This equation can be checked by differentiating \eq{eqmotion} with
respect to $\tau$.  For the linear matching \eq{linear} the proper
acceleration is given by,
\begin{equation}
\ddot{x} =-\frac{1}{2}\frac{(d-3)x^{d-2}-\Omega}{(x^{d-2}+\Omega)^2} \, .
\end{equation}
From \eq{eqmotion}  write down an ``energy'' equation:
\begin{equation}
E \equiv \frac{\zeta^2-1}{2}=\frac{1}{2}\dot{x}^2+\Phi(x) \,
\end{equation}
For different values of $E$ we analyse various solutions to the
equations of motion for radially moving test particles.  The potential
takes its maximum value $\Phi_{max}=0$ at $r=0, \infty$ and, for
$M>M_{c}$, its minimum value will be $\Phi_{min}<-\frac{1}{2}$. The
different solutions discussed below are shown as curves in
Fig. \ref{Penrose}.

\begin{enumerate}

\item For $E=0$ the particle has zero velocity at $r=0$ and
  $r=\infty$. For the linear matching the particle will start in
  region I with a non-zero velocity and cross the horizons into
  regions II and III in a finite proper time. The particle will then
  reach the centre of the black hole where it feels a repulsive force
  with a strength of $1/(2r_s\Omega)$. This force will bounce the
  particle back into regions IV and V where it will escape to
  infinity.

\item For $E>0$ the motion of the particle will be unbounded since it
  has a non-zero velocity at all points in spacetime. Starting from
  region I the particle will again move to the centre of the black
  hole crossing both horizons in a finite time $\tau$. But at $r=0$
  the particles energy will be enough to overcome the repulsive force
  and will pass through the centre of the black hole into regions IV
  and V where it will escape to infinity.

\item For $-0.5<E<0$ the particle starts with zero velocity in region
  I and continues to move into regions II and then into III where it
  has an inflection at $r>0$. Here the particle is bounced into
  regions IV and V. In region V it has a second inflection point at a
  radius equal to it's initial position in region I. The particle's
  motion is therefore bounded moving in and out of the black hole into
  different regions of spacetime.

\item For $\Phi_{min}<E<-0.5$ the particle's motion is bound to region
  II in which it has two inflection points which it moves between
  eternally.

\end{enumerate}
 
A discussion of causal structures, taking into account
the time-dependent evaporation effects of asymptotically safe black holes, 
will be summarised elsewhere.

\subsection{Role of space-time dimensionality}

It is interesting to summarize our results in view of their dependence on the space-time dimensionality, and to compare with earlier findings in four dimensions by Bonanno and Reuter  \cite{Bonanno:2000ep,Bonanno:2006eu}. 

In \cite{Bonanno:2000ep,Bonanno:2006eu}, RG improved black holes in four dimensions have been analysed using the explicit RG running \eq{G(r)2} using  \eq{linear}, \eq{Ddef} and interpolations thereof, leading to the existence of a smallest black hole whose mass $M_c$ is determined by the RG parameter $\omega$. 
We have added to this the following results. (i) 
Without specifying the explicit RG running of
Newton's coupling we have established that quantum gravity corrections imply
the existence of a smallest black hole with critical mass $M_c$, as
long as the short distance behaviour is governed by a fixed point, see
\eq{Mc_implicit}. (ii) Quantitatively, this result is largely independent of the details
of the scale matching for $k=k(r)$, which is established using the general class of matching conditions \eq{da}, and provided the short distance index 
satisfies the bound $\gamma\ge \gamma_c$ which holds for all physically motivated choices 
see \eq{gammac}. (iii) Most importantly, we have shown that 
this pattern holds true for general dimension.  In hindsight, the reason for this is that 
in fixed point gravity the graviton anomalous dimension becomes increasingly large with increasing
space-time dimensionality. Because of \eq{Mc_implicit}, the RG running of Newton's coupling can 
successfully suppress the small-$r$ singularity induced by potential term in $f(r)$.
(iv) For general space-time dimension,  the curvature singularity of the RG improved black hole is either absent or significantly reduced, compared to the classical singularity. Geodesics of the RG improved black hole space-time, for all dimensions considered, do not terminate at the curvature singularity unlike those of classical $d$-dimensional Schwarzschild black holes. This result highlights that the reduction (or absence) of curvature singularities as implied here leads to a qualitative change of the space-time structure as opposed to the classical Schwarzschild black hole, for all dimensions. 
Finally, (v)  the non-analytic threshold behaviour of low-mass black holes \eq{sqrt} for small $M-M_c$ is universal with
\begin{eqnarray}r_s(M)-r_s(M_c)&\propto& \sqrt{M-M_c}\\
T&\propto& \sqrt{M-M_c}\end{eqnarray} and independent of the dimensionality. 

In summary, we have established that the space-time dimensionality has only a small quantitative impact on the structure of RG improved black holes on all accounts addressed here. An underlying RG fixed point implies a smallest black hole whose mass $M_c$ is determined by the RG equations for gravity. Quantitatively, the main difference with increasing dimension is that the cross-over from perturbative to fixed point scaling happens in a narrower momentum-scale window.

\section{Black hole production}\label{production}

In this section, we  apply our results to the
production of mini-black holes in higher-dimensional particle physics
models of TeV scale quantum gravity.

\subsection{Large extra dimensions}
The scenario of large extra dimensions assumes that gravity propagates
in $d=4+n$ dimensions, whereas matter fields are confined to a
four-dimensional brane \cite{ArkaniHamed:1998rs,ArkaniHamed:1998}. The $n$ extra dimensions are
compactified with compactification radius $L$. For simplicity, we
assume that all radii are of the same order of magnitude, which can be
relaxed if required. The presence of extra dimensions allows for a
fundamental $d$-dimensional Planck scale $M_D$ of the order of the
electroweak scale $ \sim 1$TeV. The relationship between the effective
4-dimensional Planck scale $M_{\rm Pl}$ and the $d$-dimensional Planck
scale is given by 
\begin{equation}
 M_{\rm Pl}^2 \approx M_D^2 (M_D \,L)^n\,.  
\end{equation}
Furthermore, we require the scale-separation $M_D \,L\gg 1$ to achieve
a low fundamental quantum gravity scale. This implies that the length
scale $L$ at which the extra dimensions become visible is much larger
than the fundamental length scale $1/M_D$ at which the quantum gravity
effects become important. Consequently, at energy scales $E\approx
M_D$, the full $d$-dimensional space-time is accessible to gravity,
and our previous findings are applicable.

\subsection{Production cross section}
Here, we apply our results \cite{erg2008,MScKevin,MScAarti} to the production cross section for
mini-black holes at particle colliders. In these models, the elastic
black hole production cross section for parton-parton scattering at
trans-Planckian center-of-mass energies $\sqrt{s}\gg M_D$ is
semi-classical, provided curvature effects are small
\cite{Dimopoulos:2001hw,Giddings:2001bu,Banks:1999gd,Hsu:2002bd,Eardley:2002re}. Then, on the parton level,
the geometric cross section reads
\begin{equation}\label{sc}
\hat\sigma_{\rm cl}(s)\approx \pi\,r_{\rm cl}^2(M_{\rm phys}=\sqrt{s})\,\theta(\sqrt{s}-M_{\rm min})\,,
\end{equation}
with the physical mass replaced by the center-of-mass energy
$\sqrt{s}$. There are formation factor corrections to \eq{sc} which
have been identified in the literature, taking into account
inefficiencies in the production process (see \cite{Kanti:2004nr,Webber:2005qa,Giddings:2007nr} for
reviews). Those have not been written out explicitly as they are
irrelevant to our reasoning. For phenomenological applications, it is
often assumed that the minimal mass $M_{\rm min}$ is of the order of a
few $M_D$, limiting the regime where the semi-classical theory is
applicable.

Our study adds two elements to the picture.  The first one relates to
the threshold mass, indicating that $M_{\rm min}$ may in fact be
lower, possibly as low as the renormalised Planck mass
\begin{equation} M_{\rm min}=M_c\,.  
\end{equation}
This is a direct consequence of
the RG running of the gravitational coupling, with $M_c$ relating to
the critical physical mass (defined as in \eq{M-def}), thereby marking
a strict lower limit for the present scenario. Consequently, the RG
improved set-up has a larger domain of validity due to the weakening
of gravity at shorter distances, equally reflected in the boundedness
of the associated Bekenstein-Hawking temperature, see Sec.~\ref{t}.

\begin{figure}[t]
\includegraphics[width=.95\hsize]{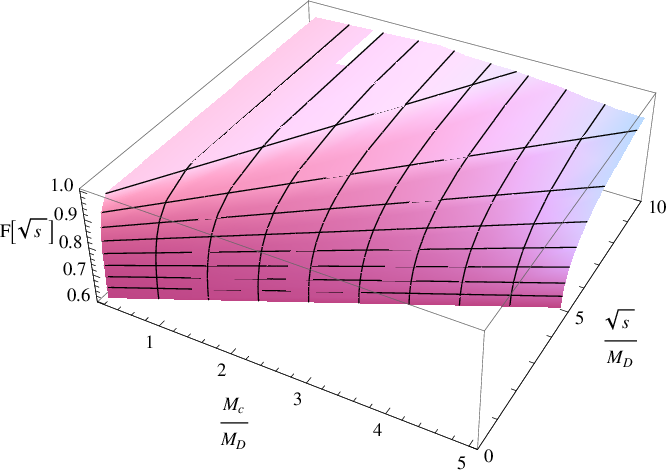} 
\caption{\label{Fs} The gravitational form factor $F(\sqrt{s})$ 
with parameter $\gamma=\gamma_{\rm dS}$ with $n=4$ extra dimensions.}
\end{figure}

The second modification takes the quantum gravity-induced reduction of
the event horizon into account, replacing $r_{\rm cl}$ by $r_s$ in
\eq{sc}. This can be written in terms of a form factor, replacing
\eq{sc} by $\hat\sigma=\hat\sigma_{\rm cl}\cdot F(\sqrt{s})$ with
\begin{equation}\label{F}
F(\sqrt{s})=\left.\left(\frac{r_s}{r_{\rm cl}}\right)^2\right|_{M_{\rm phys}=\sqrt{s}}\,,
\end{equation}
see  Fig.~\ref{Fs}. We conclude that the RG improved production 
cross section is reduced with respect to the semi-classical one,
already in the regime where the semi-classical approximation is
applicable, see Fig.~\ref{Fs}.  The quantitative impact of these effects on mini-black
holes production at colliders, eg.~the LHC, is evaluated in \cite{FHL}.

\subsection{Trans-Planckian region}\label{TP}
Next, we implement our RG improvement directly on the level of the
classical Schwarzschild radius rather than on the level of the
underlying black hole metric.  To that end, we interpret the energy
dependence of the form factor in the production cross section as
originating from an effective energy dependence of Newton's
coupling. The latter enters the classical event horizon as
\begin{eqnarray} \label{classicalBH}
r_{\rm cl}(\sqrt{s})&=& \frac{1}{\sqrt{\pi}}\left(\frac{8\,\Gamma(\frac{d-1}{2})}{d-2}\right)^{\frac{1}{d-3}}
\left(G_N\,\sqrt{s}\right)^{\frac{1}{d-3}}\quad\quad
\end{eqnarray} 
where the substitution $M_{\rm phys}=\sqrt{s}$ has already been
executed.  Under the assumption that the functional dependence of
\eq{classicalBH} on the gravitational coupling $G_N$ remains unchanged
once quantum corrections are taken into account, we can interpret the
non-trivial energy-dependence of $r_s(\sqrt{s})$ to originate from
\eq{classicalBH} via the energy-dependence of Newton's
coupling. Substituting $G_N\leftrightarrow G(\sqrt{s})$, we find
\begin{equation}\label{match-s}
{G(\sqrt{s})}={G_N}\,\left[F(\sqrt{s})\right]^\0{d-3}{2}\,.
\end{equation} 
Using the non-perturbative form factor $F(\sqrt{s})$, we conclude that
$G(\sqrt{s})$ displays a threshold behaviour, starting off at the
black hole formation threshold $\sqrt{s}\approx M_c$, and increasing
asymptotically towards $G_N$ with increasing
$\sqrt{s}$. Trans-Planckian scattering in gravity becomes classical:
$G(\sqrt{s})$ approaches $G_N$ with increasing center-of-mass energy
$\sqrt{s}\gg M_D$ and the production cross section reduces to the
geometrical one. This is a consequence of quantum corrections being
suppressed for large black hole mass, see Sec.~\ref{pt}, and thus for
large $\sqrt{s}$.

With these results at hand, we can now turn the 
argument around and identify the matching $k=k(\sqrt{s})$ which
reproduces \eq{match-s} from the renormalisation group running of
$G(k)$. To leading order in $M_D/\sqrt{s}\ll 1$, the matching
\begin{equation}\label{improved}
G_N\to G(k)\quad{\rm with}\quad k\propto 
M_D \left(\frac{M_D}{\sqrt{s}}\right)^{1/(d-3)}
 \end{equation}
in \eq{classicalBH} --- together with $G(k)$ from the renormalisation
group, see Sec.~\ref{RG} --- reproduces the form factor \eq{F} and the
semi-classical limit.
The result \eq{improved} highlights a duality between the regime of
large center-of-mass energy $\sqrt{s}/M_D\gg 1$ of a gravitational
scattering process, and the low-momentum behaviour $(k/M_D)^{d-3}
\propto M_D/\sqrt{s}\ll 1$ of the running coupling $G(k)$ in the `gravitational 
bound state' of a black hole.  

It would be interesting to have access to the behaviour of $G(\sqrt{s})$ at
below-threshold energies, where the energy dependence of Newton's coupling 
should be obtained from standard field theory
amplitudes for $s$-channel scattering in asymptotically safe gravity \cite{Litim:2007iu,Litim:2007ee}, 
which become 
strongly dominated
by multi-graviton exchange at Planckian energies \cite{'tHooft:1987rb}.  For recent developments
along these lines within quantum string-gravity, see \cite{Amati:2007ak,Marchesini:2008yh}.

\subsection{Semi-classical limit}
It is useful to compare our results with a related renormalisation
group study, where qualitatively different conclusions have been
reached \cite{Koch:2008zzb}. There, black hole production cross
sections are estimated from \eq{classicalBH} using the RG matching
\begin{equation}\label{sk}
G_N\to G(k)\quad{\rm with}\quad k \propto\sqrt{s}
\end{equation} 
for Newton's coupling, with $G(k)$ taken from the renormalisation
group and $k$ identified with $\sqrt{s}$, following \cite{Hewett:2007st}. 
This would be applicable if
$\sqrt{s}$ is the sole mass scale in the problem, and if $G_N$ in
\eq{classicalBH} is sensitive to the momentum transfer in the
$s$-channel. However, the matching \eq{sk} is in marked contrast to
\eq{improved}. Most importantly, with \eq{sk} no semi-classical limit
is achieved in the trans-Planckian regime, because $G(\sqrt{s})/G_N\ll
1$ becomes strongly suppressed. This conclusion is at variance with
the findings of the present paper.

The origin for this difference is traced back to the following 
observation: the RG improved Schwarzschild radius depends on several mass
scales, the Planck scale $M_D$, the black hole mass $M$ and, implicitly, 
the momentum scale $k$. Identifying both the mass $M=\sqrt{s}$ and 
the renormalisation group scale $k=\sqrt{s}$ with the
center-of-mass energy in a gravitational scattering process entangles
mass dependences with RG scale running.  In turn, the d\'etour taken
in Sec.~\ref{TP} disentangles these effects by taking into account
that the physics involves several mass scales.  This also
explains why $M_D$ enters the matching \eq{improved}, besides
$\sqrt{s}$, which is responsible for the qualitative difference with
respect to \eq{sk}.

We conclude that the set-up laid out in this work is necessary to capture
the semi-classical limit of trans-Planckian scattering. 

 \section{Discussion}\label{Discussion} 
%
How does quantum gravity modify the physics of black holes? 
We have implemented quantum corrections
 on the level of black hole metrics, replacing Newton's constant by
 a coupling which runs
under the renormalisation group equations for gravity.

If Newton's coupling weakens 
sufficiently fast towards shorter distances, it implies 
the existence of a smallest black hole of mass $M_c$.
This is the case  for all dimensions $d\ge 4$ provided quantum
gravity is asymptotically safe.
The mass scale  $M_c$ is dynamically 
generated and of the 
order of the fundamental Planck scale $M_D$.
Interestingly, a mere weakening of the gravitational coupling 
would not be enough to disallow the formation of an event horizon.

The mechanism responsible for a lower bound on black hole mass
relates with the RG scaling of the gravitational coupling at the
cross-over from perturbative to non-perturbative running.
In consequence, the underlying fixed point is not primarily 
responsible for the existence of the lower bound and alternative
UV completions may display a similar weakening down to 
length scales of the order of the Planck length.

  In the semi-classical regime 
$M_D/M\ll 1$,   corrections to the event horizon and black hole thermodynamics
remain perturbatively small, but effects become quantitatively more pronounced with
decreasing black hole mass $M$.
Once $M_D/M$ becomes of order  one, quantum corrections are more 
substantial.  The specific heat changes sign, the black hole temperature 
displays a maximum, and vanishes  with $M\to M_c$.
This supports the view that critical black holes
constitute cold, Planck-size, remnants. 

Direct implications of fixed point scaling are visible
in the short distance limit $r M_D\ll 1$.  This limit
becomes time-like rather than space-like as in classical Schwarzschild
black holes. Also, asymptotically safe black holes with $M>M_c$
always also display a Cauchy horizon besides the event
horizon. It is noteworthy that the classical curvature singularity at the origin is
significantly softened because of the fixed point, and either
disappears completely, or becomes vastly reduced.
The conformal structure of quantum black holes is
very similar to classical Reissner-Nordstr\"om black holes, including the near
horizon geometry of critical black holes which is of the AdS${}_2\times
{\rm S}^{\rm d-2}$ type. 

Our results have direct implications for the collider phenomenology of low-scale
gravity models. Interestingly, quantum corrections increase the domain of validity 
for a semiclassical description. At low center-of-mass energies, a threshold 
for black hole production is identified. At larger energies, quantum 
corrections to production cross sections lead to a new form factor. It reduces
the cross section, and reproduces the semi-classical result
in the trans-Planckian limit.  A
quantitative implementation of this scenario for mini-black hole
production is given elsewhere \cite{FHL}.  
It would be very interesting to
complement this picture by explicit  computations based on multi-graviton 
exchange at Planckian energies along the lines laid out in \cite{Litim:2007iu,Litim:2007ee}.\\

{\it Note added.---}
After completion of this work, we have been informed by B.~Koch that some
of our results \cite{erg2008,MScKevin,MScAarti,FHL} have been confirmed in the preprint \cite{Burschil:2009va}.

\bibliographystyle{plain}


\end{document}